\documentclass[prb,aps,amssymb,amsmath,superscriptaddress,twocolumn,showpacs,longbibliography]{revtex4-2}

\usepackage{graphicx}
\usepackage{dcolumn}
\usepackage{bm}
\usepackage{epstopdf, color}

\usepackage[utf8]{inputenc}
\usepackage{babel}
\usepackage{array}
\usepackage{amsthm}
\usepackage{amsfonts}
\usepackage{amsmath}
\usepackage{dsfont}
\usepackage{mathrsfs}
\usepackage{amssymb}
\usepackage{hyperref}
\usepackage{slashed}

\usepackage{tikz}
\usepackage{tikz-feynman}
\tikzfeynmanset{compat=1.1.0}
\usetikzlibrary{shapes}
\usetikzlibrary{positioning}

\tikzset{vertex/.style={circle,fill=black,inner sep=2pt},
	ctVertex/.style={diamond,fill=black,inner sep=2pt},
    CtVertex/.style={diamond,fill=black,inner sep=3pt},
	sqVertex/.style={rectangle,fill=black,inner sep=3pt},
	bigVertex/.style={circle,fill=black,inner sep=4pt},
	E/.append style={fill=white,draw},
    F/.append style={fill=gray,draw},
	probeEP/.style={circle,fill=black,draw,inner sep=2pt,
		prefix after command= {\pgfextra{\tikzset{every pin/.style = {pin edge={decorate,decoration={snake,amplitude=2pt,segment length =4pt}}}}}}
	},
	bareProbeEP/.style={rectangle,fill=black,draw,inner sep=3pt,
		prefix after command= {\pgfextra{\tikzset{every pin/.style = {pin edge={decorate,decoration={snake,amplitude=2pt,segment length =4pt}}}}}}
	},
	nuEP/.style={circle,fill=white,draw, inner sep=2pt},
	linelabel/.style={sloped,above,very near start, inner sep=1pt,execute at begin node=$\scriptstyle,execute at end node=$},
	baseline=(current  bounding  box.center),doubled/.style={double distance= 1pt,line width=1.5pt}
}

\newtheorem{theorem}{Theorem}[section]

\newtheorem{proposition}[theorem]{Proposition}

\newtheorem{observation}{Remark}

\def\a{\ensuremath\alpha}
\def\b{\ensuremath\beta}
\def\v{\ensuremath\vec}

\def\th{\ensuremath\theta}
\def\vth{\ensuremath\vartheta}
\def\o{\ensuremath\omega}
\def\r{\ensuremath\rho}
\def\s{\ensuremath\sigma}
\def\e{\ensuremath\epsilon}
\def\ve{\ensuremath\varepsilon}
\def\vphi{\ensuremath\varphi}
\def\hpsi{\ensuremath\hat{\psi}}

\def\de{\ensuremath\partial}

\def\mc{\ensuremath\mathcal}

\def\L{\ensuremath\Lambda}

\def\mb{\ensuremath\mathbf}

\def\dg{\ensuremath\dagger}
\def\mf{\ensuremath\mathfrak}

\begin{document}
\title{Universality of the topological phase transition in the interacting Haldane model}
\author{Simone Fabbri}%
\affiliation{SISSA, Mathematics Area, Via Bonomea 265, 34136 Trieste, Italy}
\author{Alessandro Giuliani}
\affiliation{Roma Tre University, Largo S. Leonardo Murialdo 1, 00146 Rome, Italy}
\author{Robin Reuvers}
\affiliation{Roma Tre University, Largo S. Leonardo Murialdo 1, 00146 Rome, Italy}

\begin{abstract} 
The Haldane model is a standard tight-binding model describing electrons hopping on a hexagonal lattice subject to a transverse, dipolar magnetic field. We consider its interacting 
version for values of the interaction strength that are small compared to the bandwidth. We study the critical case at the transition between the trivial and the `topological' insulating phases, and we rigorously establish that the transverse conductivity on the dressed critical line is quantized at a half-integer multiple of $e^2/h$: this is the average of the integer values of the Hall conductivity in the insulating phases on either side of the dressed critical line. Together with previous results, this fully characterizes the nature of the phase transition between different Hall plateaus and proves its universality with respect to many-body interactions. 
The proof is based on a combination of constructive renormalization group methods and exact lattice Ward identities. 
\end{abstract}

\maketitle


\section{Introduction}

One of the central questions of solid state physics is the effect of disorder and interactions on quantum transport coefficients. 
A particularly interesting  problem is to understand the dependence, if any, of the transverse conductivity of two-dimensional electron systems subject to an external, transverse magnetic field, on disorder and interactions. It is very well known \cite{TKNdN82,ASS83,BvESB94,AG98} 
that, in the independent electron approximation, the transverse conductivity of 2D lattice electron systems with or without disorder, and Fermi energy lying in a spectral or in a mobility 
gap, is quantized in integer multiples of $e^2/h$, a phenomenon known as the integer quantum Hall effect. In this setting, quantization follows from the observation that the 
Kubo conductivity is proportional to a geometrical index, the first Chern number associated with the `Bloch bundle', or the Fredholm index of an appropriate pair of Fermi projectors
\cite{ASS83, BvESB94, T94}. 

For interacting systems, quantization in integer multiples of $e^2/h$ for {\it gapped} many-fermion systems 
follows from the interpretation of the Kubo conductivity in terms of a many-body geometric index \cite{AS85,NTW85,HM15}. 
See also \cite{Gu11,ZZ22} and references therein for the definition of a topological invariant in terms of interacting Green functions.
These approaches require that the 
interacting system displays a finite spectral gap in the thermodynamic limit, which can be typically proved only for weak perturbations of gapped independent electron systems, 
where `weak' means here that the interaction strength is much smaller than the non-interacting gap \cite{dRS19,H19}. 

On the other hand, a fundamental understanding of interaction effects on the transverse conductivity in systems that, in the absence of interactions, are gapless, is 
extremely challenging and, in most respects, still open. Two common and important settings where this question is relevant are: the fractional quantum Hall effect, which concerns electron systems subject to an external magnetic field at special fillings such that a gap is expected to open thanks to the interaction (mass generation at fractional fillings, a phenomenon that 
is mostly unexplained at a fundamental, microscopic level); and the critical phase corresponding to the transition from one integer quantum Hall plateau to another. 

In this paper, we investigate the nature of the `topological transition' from the normal insulating phase to a non-trivial quantum Hall phase in a specific class of 2D interacting 
electron systems, characterized by a critical semimetallic behavior, which is the generic one on the transition line separating two distinct topological phases, both in two and three dimensions \cite{AMV18, HK10}.
The approach we follow
is not based on the introduction and use of geometrical indices or topological invariants. It is unclear whether this is at all possible for interacting semimetallic systems (see \cite{Th24} for a topological interpretation of {\it{non-interacting}} semimetals).
Therefore, rather than characterizing the Hall conductivity in terms of a geometrical index, we use a strategy that 
combines the use of Ward identities and Schwinger--Dyson equations within a constructive, rigorous renormalization group (RG) scheme (for an alternative 
RG approach to topological phase transitions, see also \cite{MCC19}). The fact that quantization of the transport coefficients can be inferred from Ward identities and Schwinger--Dyson 
equations is not new \cite{CH85,IM86}, and is related to the non-renormalization property of quantum anomalies \cite{FST96, Vo03, Wi96}, as stated e.g.\ in the Adler--Bardeen theorem \cite{AB69,M20}. 

Implementing 
these ideas within a constructive RG scheme is relatively new and, in our view, important, in that it allows one to unambiguously prove the universality, or non-universality, of transport 
coefficients, by fully taking into account finite effects due to irrelevant terms in the microscopic Hamiltonian: note that these are very difficult, if not impossible, to take into account within
formal schemes based on an effective field theory description of the system, or on standard perturbative treatments. A remarkable case in which neglecting irrelevant terms associated 
with lattice effects leads to wrong predictions is that of the optical conductivity of graphene with short-range or Coulomb interactions, where different studies of effective models of interacting graphene based on 
ultraviolet-regularized interacting Dirac fermions led 
to contradictory results, in disagreement with the experiments \cite{M08,HJV08,JVH10,HJVC08,SS09}: in the case of short-range interactions, the use of constructive RG 
methods allowed to resolve these ambiguities and to rigorously prove the universality of the optical conductivity \cite{GMP11,GMP12}. 
These methods have also been 
used to rigorously prove the universality of transport coefficients of several other interacting Fermi systems in one, two and three dimensions, including the Drude 
weight of non-integrable quantum spin chains \cite{BM11}, the longitudinal conductivity of the Haldane--Hubbard model on the critical line \cite{GJMP16,GMP20}, and 
the condensed matter analogue of the chiral anomaly in Weyl semimetals \cite{GMP21}. 

A case that remained elusive so far is that of the transverse conductivity in semimetallic critical phases, such as those at the transition between different quantum Hall 
phases in 2D interacting electron systems on the hexagonal lattice (recall that systems with hexagonal symmetry generically display a semimetallic critical behavior \cite{FW12}). The 
expectation, based on computations performed in non-interacting systems of Dirac fermions, is that in such a setting the critical transverse conductivity is quantized in a half-integer multiple of $e^2/h$, equal to the average of the two integer multiples 
displayed on the two different sides of critical state, see \cite{LFSG94} and 
\cite[Eq.(340)]{F24}. It is then argued that interactions cannot change this semi-integer value, because they are either explicitly irrelevant or marginally irrelevant as in the special case of Coulomb interactions; 
however, as discussed above in the context of the optical conductivity of graphene, this argument is inconclusive, because irrelevant terms can in general modify the values of finite quantities, including the conductivity, unless they are protected by symmetries. In this paper, we rigorously prove the quantization of the critical transverse conductivity in semi-integer multiples of $e^2/h$ in the setting of the Haldane--Hubbard model, 
i.e., the Haldane model \cite{H88} perturbed by a generalized Hubbard interaction, 
which we started to investigate in a series of previous papers \cite{GJMP16,GMP17,G20,GMP20}. The restriction to this specific setting is done just for technical simplicity and not for any physically compelling reason. We expect that our proof extends to a wider class of 2D interacting Fermi systems with critical semimetallic behavior, but we postpone such a generalization to a future publication.

\medskip

\begin{figure}[h]
    \centering
    \includegraphics[width=0.5\textwidth]{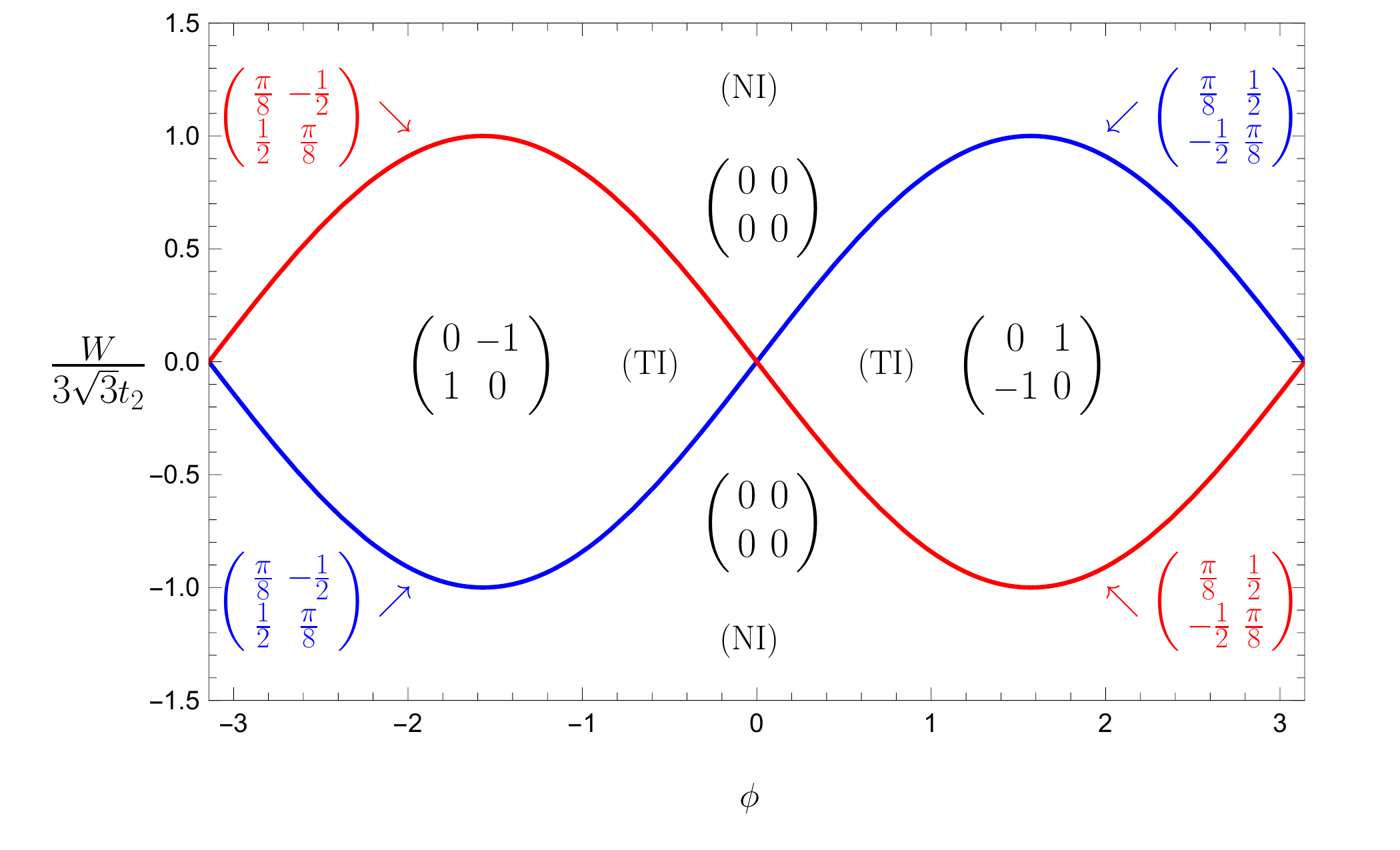}
    \caption{The conductivity matrix $\sigma$ in units of $\frac{e^2}{h}$ over the whole phase diagram (critical lines included). At the graphene points $(\phi,W)=(0,0), (\pi,0)$, the conductivity matrix is $\pi/4$ times the identity. 
    }
    \label{fig1}
\end{figure}

We recall that the Haldane model, in its non-interacting version, describes tight-binding electrons on the honeycomb lattice, subject to a transverse, dipolar magnetic field with zero net flux through the unit cell. The electrons can hop between nearest sites, with hopping strength $t_1$, and next-to-nearest sites, with hopping strength $t_2e^{i\phi}$ or $t_2e^{-i\phi}$, depending on the orientation of the next-to-nearest neighbor hopping (see Fig. \ref{fig_honeycomb}), where $\phi$ represents the line integral between the two points of the vector potential generating the magnetic field. The electrons are also subject to a local staggered potential, which takes values $+W$ and $-W$ on the even and odd sublattices of the honeycomb lattice, respectively. Assuming that $0<t_2<t_1/3$, for generic values of $\phi,W$ the valence and conduction bands are 
separated by a spectral gap. However, there are two critical curves in the $(\phi,W)$ plane, $W=\pm 3\sqrt{3}t_2\sin\phi$, at which the two bands touch: 
they divide the plane $(\phi,W)$ in four disconnected regions (see Fig. \ref{fig1}), where the energy spectrum has a non-vanishing gap and the system exhibits an integer quantum Hall effect. 

Two of these regions correspond to the topological insulator phase (TI), and are characterized by a non-trivial Hall conductance $\sigma_{12}=-\sigma_{21}= \pm \frac{e^2}{h}$, while the other two correspond to the normal insulator phase (NI), with vanishing Hall conductance. On the two critical lines, $W=\pm 3\sqrt{3}t_2\sin\phi$, the bands touch at a single point in the Brillouin zone, called the Fermi point (whose location depends on the choice of the sign in the equation of the critical line), 
around which the effective dispersion relation is approximately linear. The graphene points $(\phi,W)=(0,0), (\pi,0)$ 
are special, in that the bands touch simultaneously at both Fermi points, i.e., they display two conical intersections. The critical behavior of the system, associated with the 
transition between two plateaus of integer Hall conductivity, is, therefore, of semimetallic character. At the phase transition, both the longitudinal and transverse conductivities are quantized and non-trivial: in particular, the critical Hall conductivity turns out to be quantized at the half-integer multiple of $e^2/h$ equal to the average of the Hall conductivities on the two sides of the critical line. The computation of the critical conductivity matrix in the Haldane model is very amusing and instructive, but we could not find it in the literature: the closest computation we found is the one in \cite[Appendix 1]{LFSG94}, which, however, is based on a relativistic, linear approximation of the Hamiltonian around the Fermi points. In the lattice setting, the result is implicitly stated in Haldane's original paper, see \cite[p.2017, column 2]{H88}: roughly speaking, the argument is that at the transition one of the two Weyl components of the Dirac fermion is `heavy' and breaks time reversal symmetry, while the other is massless; the Berry curvature is not distributed uniformly over the Brillouin zone but is highly concentrated near the band crossings: so when the mass of one Weyl fermion vanishes the other still has 1/2 of the Chern number, which implies that at the transition $\sigma_{12}=\pm\tfrac12\tfrac{e^2}h$. 
A consistent computation for the full model, substantiating this argument, and performed without neglecting non-linear lattice effects, is presented in Section \ref{section_sigma_free} below. 

\medskip

The  interacting version of the Haldane model we consider is obtained by adding a generalized, finite-range, Hubbard interaction of strength $U$. The chemical potential is fixed 
in the middle of the gap between the valence and conduction bands: at criticality, it is tuned precisely to the (renormalized) energy at the 
Fermi points. In a series of previous works, we 
proved the analyticity of the ground state static correlation functions, we computed the Kubo conductivity and proved its universality, i.e., its independence from the interaction strength, 
in various regimes. The easiest case to handle is when $(\phi,W)$ is fixed away from the non-interacting critical lines and the interaction strength $U$ is sufficiently small as compared 
to the unperturbed gap: in this case, `naive' perturbation theory converges and a combined use of Ward identities and Schwinger--Dyson equations, in the spirit of \cite{CH85,IM86}, allowed us to prove the quantization of the Hall 
conductivity \cite{GMP17,G20}. In order to extend the result to all non-critical values of the parameters, in particular to situations where $U$ is small compared to the bandwidth but in general larger than the unperturbed gap, 
one needs to apply a multiscale RG scheme, with which we derived the equation for the dressed critical lines and extended the universality of the transverse conductivity to 
all values of $(\phi,W)$ outside the dressed critical lines \cite{GMP20}. In this setting, we also succeeded in proving the universality of the longitudinal conductivity {\it on} the dressed critical lines
\cite{GJMP16}. 

\begin{figure}[h]
    \centering
    \includegraphics[width=0.5\textwidth]{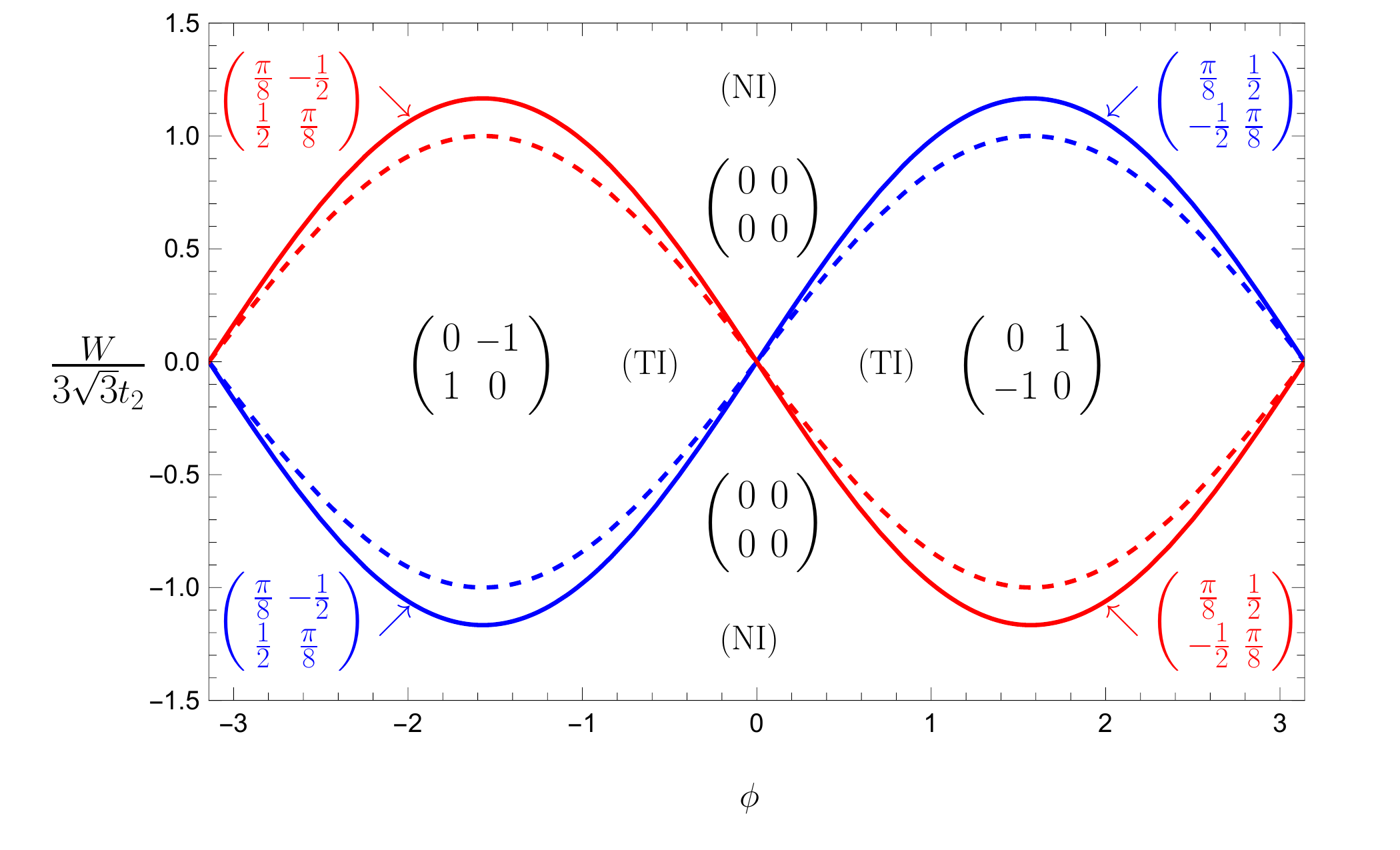}
    \caption{The phase diagram of the Haldane--Hubbard model with the corresponding conductivity matrix. The effect of the interaction shows in the renormalization of the critical curves (solid lines) compared to their non-interacting counterparts (dashed lines). According to \cite{GMP20}, the critical curves are continuously differentiable in $\phi$ and $O(U)$-close to the non-interacting ones (see \cite{GJMP16} for the leading-order computation). The conductivity matrix on and away from the dressed critical lines is the same as for the non-interacting model and, in particular, is $U$-independent.}
    \label{fig2}
\end{figure}

As mentioned above, the case of the transverse conductivity on the dressed critical lines remained elusive. The reason why the critical transverse conductivity is more 
difficult to compute than the longitudinal one, is that while the latter is dominated by relativistic contributions (the irrelevant lattice contributions being zero by parity), the critical Hall 
conductivity is dominated by irrelevant contributions, i.e., by quasi-momenta away from the Fermi points. A priori, it is unclear how to evaluate them effectively in a multiscale 
computation: constructive RG typically allows one to isolate explicit, dominant, relativistic contributions from the subleading, irrelevant ones, which are finite and equal to the {\it convergent} sum of infinitely many Feynman diagrams, but there is no simple way to evaluate them explicitly. The key to the solution, described below, is the comparison of the critical value of the transverse 
conductivity with the arithmetic average of its values at a distance $\epsilon$ inside and outside the curve: the difference between the critical value and such an average is now dominated by explicit relativistic contributions, which are shown to be zero; the irrelevant contributions to the difference are also zero, in the limit of sending the regularization parameter $\epsilon$ to zero (see the discussion after the statement of the main theorem in Section \ref{section_main} below for a more detailed description of this strategy). The fact that the arithmetic, rather than some other weighted, average is the correct one to use comes from an emergent parity symmetry close to the transition line, as discussed in Remark \ref{remark:1} below.
The resulting interacting topological phase diagram is shown in Fig.\ref{fig2}, which summarizes the findings of this and previous papers, and fully determines the 
conductivity matrix of the model for all possible choices of $(\phi,W)$, provided that $U$ is sufficiently small compared to the bandwidth. 

This paper is organized as follows. In Sec.\ref{section_Haldane}, we define the 
model and the main observables to be analyzed in this work. In 
Sec.\ref{section_main} we state our main 
result, which is summarized in Theorem \ref{thm_main}. In 
Sec.\ref{section_sigma_free}, we compute the transverse conductivity in the 
non-interacting Haldane model, while in 
Sec.\ref{section_univ} we give the proof for the interacting case, under the 
assumption that the Kubo conductivity, as computed from the RG multiscale 
expansion, can be decomposed as a sum of two terms: a `relativistic' contribution (see Eqs.\eqref{eq_15}-\eqref{eq_16}), which is dimensionally logarithmically divergent at criticality (but vanishing by parity reasons), plus a finite contribution from the irrelevant terms, which is more regular, i.e., continuous in the infrared cutoff (so that the difference between its value on the critical line and the average of its values on the two sides of the line vanishes as the infrared regularization parameter is sent to zero). 
The proof of the validity of this decomposition, 
which is summarized in Prop.\ref{proposition_main}, relies on the RG strategy developed in \cite{GJMP16, GMP20} and is discussed in Sec.\ref{section_proof}. 

\section{The model and the main results}
\label{section_Haldane}

\subsection{Set-up}

\begin{figure}[h]
    \centering
    \includegraphics[width=0.5\textwidth]{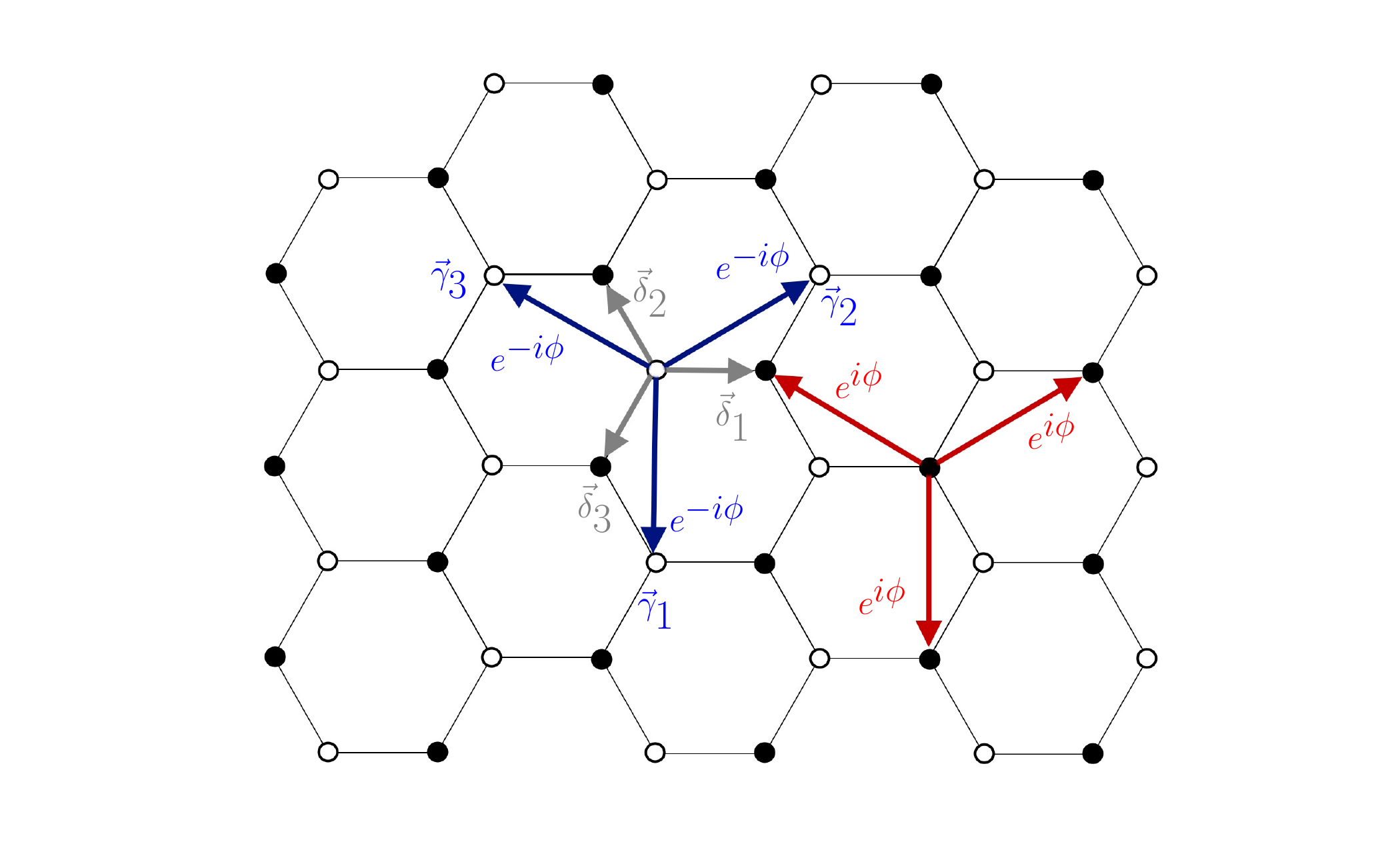}
    \caption{The honeycomb lattice of the Haldane model, with the white (resp.\ black) points corresponding to the sublattice $\Lambda$ (resp.\ $\Lambda'$). The nearest-neighbor vectors $\v{\delta}_1, \v{\delta}_2,\v{\delta_3}$ and next-to-nearest neighbor vectors $\v{\gamma}_1,\v{\gamma}_2,\v{\gamma}_3$ are also represented. For the latter, the phases associated to the hoppings from a white (resp.\ black) site to their next-to-nearest neighbor are shown in blue (resp.\ red); the hoppings in the opposite directions have complex conjugate phases.}
    \label{fig_honeycomb}
\end{figure}

We shall think of the hexagonal lattice, which our
model is defined on, as the superposition of 
two triangular lattices $\L$ and $\L'$, shifted with respect to one another. We let $\Lambda=\cup_{n\in\mathbb Z^2}\{n_1\v{\ell}_1+n_2\vec\ell_2\}$ be the infinite triangular lattice generated by the basis vectors
\[ 
\v{\ell}_1= \frac{1}{2}(3, -\sqrt{3}), \hspace{1cm} \v{\ell}_2= \frac{1}{2}(3, \sqrt{3}),
\]
and $\L'= \L +(1,0)$. We then label each point of the hexagonal lattice by a pair $(\v{x},\r)$ with $\v{x}\in\Lambda$ and $\r\in\{1,2\}$, with the understanding that $(\v{x}, 1)$ corresponds to $\v{x}\in\L$ and $(\v{x}, 2)$ to $\v{x}+(1,0)\in\L'$ (see Fig. \ref{fig_honeycomb}). We denote by $\psi^\dagger_{\v{x},\rho}$
and $\psi_{\v{x},\rho}$ the fermionic creation and annihilation operators at site $(\v{x},\rho)$, respectively, and by $\psi^\dagger_{\v{x}}$ and $\psi_{\v{x}}$ the corresponding $2$-component spinors. 

Given a positive integer $L$, we study the model on the discrete torus $\L_L=\L/L\L$; the box size $L$ acts as an infrared cutoff that will eventually be sent to infinity. We use the following convention for the Fourier transform of the fermionic operators:
\[
\hat{\psi}_{\v{k}}= \sum_{\v{x}\in\L_L}e^{i\v{k}\cdot\v{x}}\psi_{\v{x}}, \hspace{0,5cm} {\psi}_{\v{x}}= \frac{1}{L^2}\sum_{\v{k}\in\mc{B}_L}e^{-i\v{k}\cdot\v{x}}\hat\psi_{\v{k}},
\]
with $\v{k}$ belonging to the discretized Brillouin zone
\begin{equation}\label{mBL}
 \mc{B}_L:= \left\{\v{k}=\tfrac{n_1}{L}\v{G}_1+ \tfrac{n_2}{L}\v{G}_2: n_i\in \mathds{Z}\cap\left(-\tfrac{L}{2},\tfrac{L}{2}\right]  \right\},
\end{equation}
which is 
defined in terms of the vectors $\v{G}_1,\v{G}_2$ specified by the condition $\v{G}_i\cdot\v{\ell}_j= 2\pi \delta_{ij}$. $\mathcal B_L$ should also be thought of as a discrete torus, i.e., the sum of any two vectors in $\mathcal B_L$ should be identified with an element of $\mathcal B_L$ modulo integer translations by $\vec G_1$ and/or $\vec G_2$. We let $\mc{B}\equiv \mathcal B_\infty$ be the infinite-volume Brillouin zone.

The non-interacting, finite-volume Hamiltonian with periodic boundary conditions is \cite{H88}
\[
\begin{split}
    &\mc{H}_{0,L}=
    -t_1\sum_{\v{x}\in\Lambda_L}\left[ \psi^{\dagger}_{\v{x},1}\left(\psi_{\v{x},2} + \psi_{\v{x}-\v{\ell}_1,2} + \psi_{\v{x}-\v{\ell}_2,2} \right) + h.c. \right]\\
    &-t_2 \sum_{\v{x}\in\Lambda_L}\sum_{\substack{\alpha=\pm\\ j=1,2,3}}\left(e^{i\alpha\phi} \psi^{\dg}_{\v{x},1}\psi_{\v{x}+\alpha\v{\gamma}_j,1} + e^{-i\alpha\phi} \psi^{\dg}_{\v{x},2}\psi_{\v{x}+\alpha\v{\gamma}_j,2} \right)\\
    &+ W\sum_{\v{x}\in\Lambda_L}\left(\psi^{\dag}_{\v{x},1}\psi_{\v{x},1} - \psi^{\dag}_{\v{x},2}\psi_{\v{x},2} \right),
\end{split}
\]
where: the first line describes nearest neighbor hopping by vectors $\vec\delta_1=(1,0)$, $\vec\delta_2=(-\tfrac12,\tfrac{\sqrt3}2)$, $\vec\delta_3=(-\tfrac12,-\tfrac{\sqrt3}2)$, see Fig.\ref{fig_honeycomb}; the second describes next-to-nearest neighbor hopping by vectors
\[
\v{\gamma}_1= \v{\ell}_1 - \v{\ell}_2,\ \v{\gamma}_2=\v{\ell}_2,\ \v{\gamma}_3= -\v{\ell}_1,
\]
see again Fig.\ref{fig_honeycomb}; and the third describes a staggered local potential, favoring occupancy of the black (resp.\  white) sites for $W>0$ (resp.\ $W<0$). We fix the hopping strengths $t_1,t_2$ once and for all; for definiteness, we assume $t_1,t_2>0$; moreover, in order to avoid having overlapping bands (see Section \ref{sec2C}), we assume that $t_2$ is not too large as compared to $t_1$, namely $t_2/t_1<1/3$. Given $t_1,t_2$ with these properties, the non-interacting Hamiltonian will be thought of as being parametrized by $W$ and $\phi$, and in order to make such dependence explicit, we shall write $\mathcal H_{0,L}=\mathcal H_{0,L}(W,\phi)$. 

The interacting Hamiltonian in the grand-canonical ensemble is 
\begin{equation}
\label{def_Hamiltonian}
    \mc{H}_L= \mc{H}_{0,L}(W,\phi) - \mu \mc{N}_L+ U\mc{V}_L,
\end{equation}
which has to be thought of as a function of the parameters $W,\phi,\mu,U$, where $\mu$ is the chemical potential and $U$ is the interaction strength; in \eqref{def_Hamiltonian}, letting $n_{\v{x},\r}= \psi^{\dg}_{\v{x},\r}\psi_{\v{x},\r}$ be the local density,
$\mc{N}_L=\sum_{\v{x}\in\Lambda_L}\sum_{\rho=1,2}n_{\vec x,\rho}$ is the total 
number operator, and $\mc{V}_L$ is a generalized Hubbard interaction, of the form 
\begin{equation}\label{defVL}
\mc{V}_L= \sum_{\v{x},\v{y}\in\L_L}\sum_{\r,\r'\in\{1,2\}} (n_{\v{x},\r} -\tfrac{1}{2})v_{\r,\r'}(\v{x}-\v{y}) (n_{\v{y},\r'} -\tfrac{1}{2} ).
\end{equation}
The potential $v_{\r,\r'}$ satisfies $v_{11}(\v{x})= v_{22}(\v{x})=v(\v{x})$, $v_{12}(\v{x})= v(\v{x}-(1,0))$, $v_{21}(\v{x})= v(\v{x}+(1,0))$, with $v$ a rotationally-invariant, finite-range potential. 
The terms $-\tfrac12$ in the definition of $\mathcal V_L$ provide a useful normalization of the chemical potential, which turns out to be convenient in the Grassmann representation of the model described below. In fact, thanks to the presence of the terms $-\tfrac12$ in \eqref{defVL}, the Grassmann counterpart of the interaction $\mathcal V_L$ is purely quartic, see the first line of \eqref{interaction}
(for a proof that the Hamiltonian and Grassmann versions of the quartic interaction differ by these $-\tfrac12$ factors, see \cite[Sect.5.1]{GMP17}, compare in particular 
\cite[eq.(5.2)]{GMP17} and \cite[eq.(5.9)]{GMP17}). 

\subsection{Correlation functions and Kubo conductivity}
\label{section_observables}

The central object of our interest is the Kubo conductivity, which is written in terms of the two-point current correlation function. We recall that the lattice current operator is defined by first promoting the Hamiltonian $\mathcal H_L$ to be $\vec A$-dependent, where $\vec A$ is an external $U(1)$ vector potential, via a replacement of the hopping parameters to $\vec A$-dependent hoppings via the Peierls substitution; and then by taking $\vec{\mathcal J}_{\vec{p}}=\delta H(\vec A)/\delta \vec A_{\vec p}\big|_{\vec A=\vec{0}}$, see \cite[Appendix A]{GJMP16}; or, equivalently, it can be defined by imposing the lattice continuity equation, see \cite[Sect.2.2]{GMP20}. The result is
\begin{equation}
\label{currentoperator}
\v{\mc{J}}_{\v{p}}= \frac{1}{L^2}\sum_{\v{k}\in\mc{B}_L} \hpsi^{\dg}_{\v{k}+\v{p}}\v{\Gamma}(\v{k},\v{p})\hpsi_{\v{k}},
\end{equation}
where $\v{p}\in \mc{D}_L:= 
\{\v{k}=\tfrac{n_1}{L}\v{G}_1+ \tfrac{n_2}{L}\v{G}_2: n_i\in \mathds{Z} \}$, and the components of the vector $\v{\Gamma}(\v{k},\v{p})$ are $2\times 2$ matrices, called the bare vertex functions, such that 
\begin{equation}\label{eqGammas}
\begin{split}
&\v{\Gamma}_{1,1}(\v{k},\v{p})= -it_2\sum_{j=1}^3\sum_{\a=\pm}\a \v{\gamma}_j\eta_{\a\v{\gamma}_j\cdot\v{p}} e^{i\a(\phi-\v{k}\cdot\v{\gamma}_j)}\\
&\v{\Gamma}_{1,2}(\v{k},\v{p})= -it_1\sum_{j=1}^3\v{\delta}_j\eta_{\v{\delta}_j\cdot\v{p}}e^{-i\v{k}\cdot(\v{\delta}_j-\v{\delta}_1)},
\end{split}\end{equation}
with $\eta_x:= (e^{-ix}-1)/(-ix)$, while $\v{\Gamma}_{2,1}(\v{k},\v{p})=-\v{\Gamma}_{1,2}(-\v{k}-\v{p},\v{p})$, and $\v{\Gamma}_{2,2}(\v{k},\v{p})= -e^{-i\v{p}\cdot\v{\delta}_1}\v{\Gamma}_{1,1}(-\v{k},-\v{p})$. 

In particular, the current operator at $\v{p}=\v{0}$ can be written as (cf. \cite[Appendix A]{GMP17})
\begin{equation}
\label{eq_4}
\v{\mc{J}}_{\v{0}}= -\frac{1}{L^2}\sum_{\v{k}\in\mc{B}_L} \tilde{\psi}^{\dg}_{\v{k}} \nabla_{\v{k}}\tilde{H}^0(\v{k}) \tilde{\psi}_{\v{k}},    
\end{equation}
where $\tilde{\psi}_{\v{k}}= \Gamma^\dg_0(\v{k})\hpsi_{\v{k}}$ and $\tilde{H}^0(\v{k})= \Gamma^\dg_0(\v{k})\hat{H}^0(\v{k}) \Gamma_0(\v{k})$. 
Here, $\Gamma_0(\v{k})= \begin{pmatrix}
1     &  0\\
0     & e^{-i k_1}\end{pmatrix}$ and 
$\hat H^0(\vec k)$ is the Bloch Hamiltonian of the non-interacting system (cf.\ \cite{H88} and \cite[Appendix B]{GMP17}):
\begin{eqnarray}
\label{eq33}
&&\hat{H}^0(\v{k})=\label{eqH0}\\
&& \left(\begin{array}{cc}
-2t_2\a_1(\v{k})\cos\phi + m(\v{k})& -t_1\Omega^*(\v{k})\\
-t_1\Omega(\v{k}) & -2t_2\a_1(\v{k})\cos\phi - m(\v{k})\\
\end{array}\right),\nonumber
\end{eqnarray}
where
\begin{alignat*}{2}
\a_1(\v{k})&= \sum_{j=1}^3\cos(\v{k}\cdot\v{\gamma}_j),\hspace{0.5cm}
&&m(\v{k})= W -2t_2\a_2(\v{k})\sin\phi,\\
\a_2(\v{k})&= \sum_{j=1}^3\sin(\v{k}\cdot\v{\gamma}_j),\hspace{0.5cm}
&&\Omega(\v{k})= 1+ e^{-i\v{k}\cdot\v{\ell}_1} + e^{-i\v{k}\cdot\v{\ell}_2}.
\end{alignat*}
In the thermodynamic and zero-temperature limits, after a Wick rotation of the time variable, the Kubo conductivity matrix of elements $\sigma_{ij}$ can be expressed in terms of Euclidean current-current correlations, see \cite[Theorem 3.1]{GMP17} for a proof of the validity of the Wick rotation for the conductivity in the off-critical case, and \cite[Sec.5]{MP22} for a proof in the critical case. One finds that
\begin{equation}
\label{def_Kubo}
\s_{ij}= \frac{1}{|\v{\ell}_1\times\v{\ell}_2|}\partial^-\hat{K}_{ij}(0),
\end{equation}
where $i,j\in\{1,2\}$, $|\v{\ell}_1\times\v{\ell}_2|=3\sqrt{3}/2$ is the unit cell area, $\partial^-\hat{K}_{ij}(0)=\lim_{p_0\to 0^-}\tfrac{\hat K_{ij}(p_0)-\hat K_{ij}(0)}{p_0}$ is the left derivative at $p_0=0$, 
and 
\begin{equation}
\label{def_K}
\begin{split}
&\hat {K}_{ij}(p_0)=\lim_{\beta\to\infty}\lim_{L\to\infty}\frac{1}{\beta L^2}\int_0^\beta dt_1\int_0^\beta dt_2\, e^{-ip_{0,\beta}(t_1-t_2)}\cdot\\
&\quad \cdot \Big[\langle T\big(\mc{J}_{\v{0},i}(t_1)\mc{J}_{\v{0},j}(t_2)\big)\rangle_{\b,L}-\langle\mc{J}_{\v{0},i}\rangle_{\b,L}\langle\mc{J}_{-\v{0},j}\rangle_{\b,L}\Big]
\end{split}
\end{equation}
is the Euclidean current-current correlation at Matsubara frequency $p_0$ in the thermodynamic and zero-temperature limits. In \eqref{def_K}, the expectation $\langle \cdot\rangle_{\beta,L}$ is computed with respect to the Gibbs measure at inverse temperature $\beta$,
\begin{equation*}\langle \cdot \rangle_{\b,L}= \frac{\mbox{Tr}\left\{e^{-\b \mc{H}_L} \cdot \right\}}{\mbox{Tr}\left\{e^{-\b \mc{H}_L} \right\}},\end{equation*} and, for any $0\le t<\beta$, $\mc{J}_{\v{p},i}(t)=e^{t \mc{H}_L} \mc{J}_{\v{p},i}e^{-t\mc{H}_L}$ is the imaginary time evolution of the current; moreover, $p_{0,\beta}=\tfrac{2\pi}{\beta}\lfloor \tfrac{\beta p_0}{2\pi}\rfloor$ is an integer multiple of $2\pi/\beta$ tending to $p_0$ as $\beta\to\infty$; finally, 
the operator $T$ inside the expectation on the RHS is the time-ordering operator, which reorders the product of the two time-dependent operators in its argument in decreasing time order.


\medskip

For later reference, we also introduce the two-point Schwinger function in the thermodynamic and zero-temperature limits (i.e., the Euclidean Green's function) 
\begin{equation}\begin{split} \hat{S}(k_0,\vec{k})&=\lim_{\beta\to\infty}\lim_{L\to\infty}\tfrac1{\beta}\int_0^\beta dt_1\int_0^\beta dt_2 \sum_{\vec x\in\Lambda_L}\cdot\\
& \cdot e^{i k_{0,\beta}(t_1-t_2)+i\vec k_L\cdot\vec x}
\langle T\big( \psi_{\v{x}}(t_1) \psi^{\dg}_{\v{0}}(t_2)\big)\rangle_{\b,L},
\end{split}\label{def_Schwinger}\end{equation}
where $k_{0,\beta}=\tfrac{2\pi}{\beta}(\lfloor \tfrac{\beta\omega}{2\pi}\rfloor+\tfrac12)$, $(\vec k_L)_i=\tfrac{2\pi}{L}\lfloor \tfrac{L k_i}{2\pi}\rfloor$,
and, again, $\psi^{(\dagger)}_{\v{x}}(t)=e^{t \mc{H}_L}\psi^{(\dagger)}_{\v{x}}e^{-t\mc{H}_L}$ indicates the time evolution of the creation or annihilation fermionic operators, while $T$ is the fermionic time ordering operator such that
\begin{equation*} T\big( \psi_{\v{x}}(t_1) \psi^{\dg}_{\v{0}}(t_2)\big)=
\begin{cases} \psi_{\v{x}}(t_1) \psi^{\dg}_{\v{0}}(t_2) &\text{if $t_1>t_2$} \\
-\psi^{\dg}_{\v{0}}(t_2)\psi_{\v{x}}(t_1) &\text{if $t_1\le t_2$}.
\end{cases}
\end{equation*}

\subsection{The non-interacting theory}\label{sec2C}

Before proceeding further, it is convenient to briefly recall a few  properties of the system at $U=0$, in which case the model is exactly solvable. The energy bands, i.e., the $\vec k$-dependent eigenvalues of the Bloch Hamiltonian \eqref{eqH0}, are
\[
    \ve_{\pm}(\v{k})= -2t_2\a_1(\v{k})\cos\phi \pm \sqrt{m(\v{k})^2 + t_1^2|\Omega(\v{k})|^2}.
\]
Under the assumption $t_2/t_1<1/3$, the two bands do not overlap, and can only touch at the two Fermi points, where  $\Omega(\vec k)$ vanishes: $\v{k}_F^{\,\pm}= \frac{2\pi}{3}\big(1,\pm\tfrac{1}{\sqrt3}\big)$.
Denoting $m_\o:=m(\v{k}_F^\o)$, with $\o\in\{+,-\}$, the two bands touch at $\v{k}_F^\o$ iff
\begin{equation}
\label{critlines}
m_\o(\phi,W)=W+ 3\sqrt{3}\o t_2\sin\phi=0.
\end{equation}
This equation defines two curves in the $(\phi,W)$-plane, which we call the \textit{critical curves} of the non-interacting theory.
We fix the chemical potential at $\mu=
-2t_2 \a_1(\v{k}_F^+)\cos\phi=3t_2\cos\phi$, so that, if $(\phi,W)$ is on the critical curves, the system is a semimetal, while, in the complement of the critical curves, it is in an insulating (gapped) phase. In that case, 
the computation of the conductivity matrix leads to the values indicated in Fig.\ \ref{fig1}: the computation in the off-critical, insulating case was discussed in \cite[Appendix B]{GMP17}; the computation of the longitudinal conductivity on the critical lines was discussed in \cite[Sect.IV]{GMP11} (at the graphene point) and in \cite[Sect.IV.B]{GJMP16} (away from the graphene point); the computation of the transverse conductivity on the criticial lines, away from the graphene points where it is trivially zero, is discussed in Section \ref{section_sigma_free} below. 

\medskip
For completeness, let us conclude this section by describing the form that the two-point Schwinger function \eqref{def_Schwinger} takes in the non-interacting case, 
\begin{equation}
\label{eq_1}
 \hat{S}^0(k_0,\vec k):=\hat S(k_0,\vec k)\big|_{U=0}=\big( -ik_0 + \hat{H}^0(\v{k}) -\mu \big)^{-1},
\end{equation}
with $\mu=3t_2\cos\phi$. Its Fourier dual is denoted by $S^0(t_1-t_2,\vec x)=
\lim_{\beta\to\infty}\lim_{L\to\infty}\langle T\big( \psi_{\v{x}}(t_1) \psi^{\dg}_{\v{0}}(t_2)\big)\rangle_{\b,L}\big|_{U=0}$, and its partial Fourier dual (with respect to the Matsubara frequency) by
$\widetilde S^0(t,\vec k):=\sum_{\vec x\in\Lambda}e^{i\vec k\cdot\vec x}
S^0(t,\vec x)$, which reads 
\begin{equation}
\tilde{S}^0(t,\v{k})=e^{-t(\hat{H}^0(\v{k})-\mu)} \left(\mathds{1}_{\{t>0\}} 
P_+(\v{k}) - \mathds{1}_{\{t\le 0\}} P_-(\v{k}) \right),
\end{equation}
where $P_{+}(\v{k})$ and $P_-(\vec k)$ are the projections over the upper and 
lower bands, respectively.

If both $m_+$ and $m_-$ are different from zero, then $\hat S^0(\mb k)$ 
is real analytic in $\mb{k}\equiv(k_0,\vec k)$ over $\mathbb{R}\times\mc{B}$, 
and its Fourier dual $S^0(\mb x)$ decays exponentially over $\mathbb R\times \Lambda$.
Conversely, on the critical lines defined by \eqref{critlines}, $\hat S^0(\mb k)$ in \eqref{eq_1} has a simple pole at $\mb{k}_F^{\o}\equiv (0,\v{k}_F^{\o})$:
\begin{equation}
\label{eq_2}
\begin{split} &\hat{S}^0\left(\mb{k}_F^{\o} +\mb{k}'\right)=\\ 
&\left(\begin{array}{cc}
-ik_0'& \frac{3}{2}t_1(ik_1'-\o k_2')\\
-\frac{3}{2}t_1(ik_1'+\o k_2') & -ik_0'\\
\end{array}\right)^{-1} \left(1+ O(|\mb{k}'|) \right), 
\end{split}
\end{equation}
as $|\mb k'|\to 0$, and, correspondingly, $S^0(\mb x)$ decays algebraically at large time-space distances, asymptotically bounded from above and below by a constant times $|\mb x|^{-2}$. 

\subsection{Main results: the Hall conductivity in the critical regime}
\label{section_main}

As anticipated in the introduction, our main result concerns the transverse conductivity $\sigma_{12}=-\sigma_{21}$, and can be stated as follows. 

\begin{theorem}
\label{thm_main}
There exists $U_0>0$, independent of $W,\phi$, 
such that, for any $|U|<U_0$, there exist two functions $\mf{d}(U,W,\phi), \mf{z}(U,W,\phi)$, analytic in $U$, vanishing at $U=0$, and continuously differentiable in $W,\phi$, 
such that, if the chemical potential $\mu$ is fixed at the value $3t_2\cos\phi - \mf{z}(U,W,\phi)$, the Hall conductivity of the interacting Haldane model \eqref{def_Hamiltonian} reads: 
\begin{equation}
\label{eq_main}
\s_{12}= \tfrac{1}{4\pi}\Big[\textup{ sgn}\left(m^R_+(U,W,\phi)\right) - \textup{ sgn}\left(m^R_-(U,W,\phi)\right)\Big],
\end{equation}
where $m^R_{\pm}(U,W,\phi)\equiv W \pm 3\sqrt{3}t_2 \sin\phi \pm \mf{d}(U,W,\phi)$, with the understanding that $\textup{sgn}(0)=0$.
\end{theorem}

As discussed in \cite{GJMP16,GMP20}, the functions $m^R_\pm$ have the meaning of dressed, renormalized, masses. The system is critical, i.e., its Euclidean correlation functions 
decay polynomially to zero at large space-time separation, iff either of the two masses vanishes. 

The proof of the theorem is constructive, that is, it is based on an algorithm allowing one to compute $U_0$, as well as the Taylor coefficients of the analytic functions $\mf{d}$ and $\mf{z}$, representing the interaction-induced shifts 
of the mass and of the chemical potential, respectively (see the discussion after eq.\eqref{refhamiltonian}, 
and \cite[Lemma 4.2]{GMP20} for additional details; see also \cite[Section III.E]{GJMP16} for the explicit lowest-order computation of $\mf{z}$ in the case of ultra-local Hubbard interaction). The value of $U_0$ provides an estimate on the strength of the interaction beyond which new physics appears. However, in this paper we do not attempt to evaluate it explicitly, because we do not expect that our proof can provide a realistic value for the transition strength. A numerically more refined scheme and a computer-assisted proof would be required for this purpose, but this goes beyond the scope of this work.

The proof of Theorem \ref{thm_main} in the off-critical case, when both $m^R_{+}(U,W,\phi)$ and $m^R_{-}(U,W,\phi)$ are different from zero, was treated in \cite{GMP20}. 

The new case proved in this paper is the critical case where either $m^R_{+}(U,W,\phi)$ or $m^R_{-}(U,W,\phi)$ vanishes (the case where both simultaneously vanish is easier, and was treated in \cite{GMP11,GJMP16}): 
this was precisely the missing case needed to complete the picture of the conductivity matrix displayed in Fig. \ref{fig2}. The reason why the computation of the critical $\s_{12}$ remained an open problem is that none of the methods used in the earlier works \cite{GJMP16,GMP20} is directly applicable in the present case. 

In \cite{GMP20}, we proved the quantization of $\sigma_{12}$ in the off-critical case via a combination of Ward identities and Schwinger--Dyson equations, which implies the vanishing of the non-universal corrections to the conductivity provided that the current-current correlations in momentum space are three times differentiable, see \cite[Sect.3.1]{GMP20}. 

Such a differentiability condition fails on the critical line, where the derivative of the Euclidean current-current correlation, $\tfrac{\partial}{\partial p_0}\hat K_{ij}(p_0)$, is dimensionally logarithmically divergent as $p_0\to 0$ (note that $\sigma_{ij}$ is proportional precisely to the right derivative at $p_0=0$ of such an a priori singular quantity). Therefore, at criticality, one needs to proceed in a different way. An effective strategy, which works well in the case of the longitudinal conductivity \cite{GMP11,GJMP16}, is to rewrite $\hat K_{ij}(p_0)$ as the sum of a singular contribution, coming from the lowest-order Feynman diagram (the `polarization bubble') with dressed, renormalized, vertex functions, and of a regular one, given by the convergent sum of all the dressed diagrams with at least one interaction vertex; note, in fact, that the quartic interaction is dimensionally irrelevant in the infrared, which induces a dimensional gain with respect to the naive power counting on all the interaction corrections beyond the dressed polarization bubble; this, in turn, implies that the such higher order interaction corrections sum up to a quantity that is continuously differentiable at $p_0=0$. On the other hand, elementary parity considerations show that the longitudinal current-current correlation, $\hat K_{ii}(p_0)$, is {\it even} in $p_0$, and so are its singular and regular parts, separately. Therefore, the derivative of the regular part of $\hat K_{ii}(p_0)$ at $p_0=0$ is readily zero, and so is its contribution to the critical longitudinal conductivity. On the other hand, the contribution to $\partial\hat K_{ii}(p_0)$ as $p_0\to 0^+$, coming from the polarization bubble can be evaluated explicitly and gives $1/8$ per Dirac cone, see \cite[Sect.IV.B]{GJMP16}. 

Unfortunately, a direct application of this strategy fails for the critical transverse conductivity. In fact, by parity reasons, it turns out that things go the other way round for the off-diagonal components of the conductivity matrix: the singular contribution from the polarization bubble vanishes and the whole contribution to $\sigma_{12}$ comes from the regular part. However, this regular part is not explicit: it is the convergent sum of infinitely many Feynman diagrams and there is no hope to compute the sum directly. 

\medskip 

 In this paper, we show how to compute the critical transverse conductivity via two different strategies, different from those of earlier papers. The first one, discussed in Section \ref{section_sigma_free}, applies to the non-interacting theory, and is based on methods closely resembling the ones used in the off-critical case (see e.g. \cite[Appendix B]{GMP17}), which admit an explicit interpretation in terms of the Berry curvature of the Bloch bundle. The second proof, discussed in Sections \ref{section_univ} and  \ref{section_proof}), is more general, valid both in the non-interacting and interacting cases. It consists in comparing the critical transverse conductivity with an appropriate average of the off-critical values of $\sigma_{12}$ inside and outside the critical curve, at a distance $\epsilon$ to be eventually sent to zero. The difference between the critical $\sigma_{12}$ and the average of its off-critical values at distance $\epsilon$ can also be decomposed in its singular and regular parts via a multiscale renormalization group computation, which can both be shown to vanish as the regularization parameter $\epsilon$ goes to zero, either via a direct computation (as far as the singular contribution from the polarization bubble is concerned) or via dimensional bounds (as far as the regular part from the higher order interaction corrections is concerned). 

\section{The critical transverse conductivity}
\subsection{The non-interacting case}
\label{section_sigma_free}

In this section we compute $\sigma_{12}$ in the critical, non-interacting case, and prove in this setting the validity of \eqref{eq_main}. 
We let $U=0$ and, without loss of generality, we focus on the critical line $m_-=0$, with $\phi\in (0,\pi)$: the cases where $m_+=0$ and/or $\phi\in(-\pi,0)$ are equivalent thanks to the symmetries of the model, see \cite[Sect.III.B]{GJMP16}.

Our starting point is Kubo's formula \eqref{def_Kubo}. 
Plugging \eqref{eq_4} into \eqref{def_K} and using Wick's rule, we find
\begin{equation}
\label{eq_5}
\begin{split}
&\hat{K}_{ij}(p_0)= - \int_{\mathds{R}}dt \int_{\mc{B}}\frac{d\v{k}}{|\mc{B}|} e^{-ip_0t} \\
&\cdot \mbox{Tr}\left\{\tilde{g}(-t,\v{k})\de_{k_i}\tilde{H}^0(\v{k})\tilde{g}(t,\v{k}) \de_{k_j}\tilde{H}^0(\v{k}) \right\}.
\end{split}
\end{equation}
Here,
\[
\tilde{g}(t,\v{k}):=e^{-t\tilde{H}^0(\v{k})} \left(\mathds{1}_{\{t>0\}} \tilde{P}_+(\v{k}) - \mathds{1}_{\{t\le 0\}} \tilde{P}_-(\v{k}) \right),
\]
and $\tilde{P}_{\pm}(\v{k})= \Gamma^\dg_0(\v{k})P_{\pm}(\v{k}) \Gamma_0(\v{k})$, with $P_{\pm}(\v{k})$ the projections on the upper and lower bands, respectively.
Plugging \eqref{eq_5} into \eqref{def_Kubo} we find that
\begin{equation}
\label{eq_6}
\begin{split}
&\s_{ij}=-\lim_{p_0\rightarrow 0^-} \int_{\mathbb{R}} dt \int_{\mc{B}} \frac{d\v{k}}{(2\pi)^2}\left(\frac{e^{-ip_0t}-1}{p_0}\right) \\
&\cdot  \mbox{Tr}\left\{\tilde{g}(-t,\v{k})\de_{k_i}\tilde{H}^0(\v{k})\tilde{g}(t,\v{k}) \de_{k_j}\tilde{H}^0(\v{k}) \right\},
\end{split}
\end{equation}
where we used the fact that $|\v{\ell}_1\times\v{\ell}_2||\mc{B}|= (2\pi)^2$. By separating the integral over $(0,+\infty)$ from the one over $(-\infty,0)$ in \eqref{eq_6} and by changing variable $t\rightarrow-t$ in the latter, we obtain
\begin{equation}
\label{eq_7}
\begin{split}
&\s_{ij} = \lim_{p_0\rightarrow 0^-}\int_0^{\infty} dt \int_{\mc{B}} \frac{d\v{k}}{(2\pi)^2} \\
&\cdot \left[\left(\frac{ e^{-ip_0t}-1}{p_0}\right) f_{ij}(t,\v{k}) + \left(\frac{ e^{ip_0t}-1}{p_0}\right) f_{ji}(t,\v{k}) \right], 
\end{split}
\end{equation}
with
\begin{equation}
\label{eq_30}
\begin{split}
&f_{ij}(t,\v{k})\\
&:=\mbox{Tr}\Big\{ e^{t\tilde{H}^0(\v{k})} \tilde{P}_-(\v{k}) \de_{k_i}\tilde{H}^0(\v{k}) e^{-t\tilde{H}^0(\v{k})} \tilde{P}_+(\v{k}) \de_{k_j}\tilde{H}^0(\v{k}) \Big\}. \end{split}
\end{equation}
We are studying the critical line $m_-=0$ and $\phi\in (0,\pi)$, which means that the energy gap only closes at the Fermi point $\v{k}_F^-$. It is therefore convenient to split the integral over momenta in \eqref{eq_7} into an integral over a small ball centered at $\v{k}_F^-$, and an integral over its complement. That is, we introduce $B_{\e}:= \left\{\v{k}\in\mc{B}: |\v{k}-\v{k}_F^-|\le \e \right\}$ with $\e \ll 1$, and write
\[
\s_{12}= \s_{12}^{(a)}+ \s_{12}^{(b)},
\]
where $\s_{12}^{(a)}$ and $\s_{12}^{(b)}$ are given by the RHS of \eqref{eq_7}, with the integral over $d\v{k}$ restricted to $B_{\e}$ and $\mc{B}\setminus B_{\e}$ respectively.

\medskip

\textit{The contribution $\s_{12}^{(a)}$}. We focus on the first term on the RHS of \eqref{eq_7}. Using \eqref{eq_30} and recalling the definition of $\Gamma_0(\vec{k})$ (see the lines after \eqref{eq_4}), we rewrite
\begin{equation}
\label{eq_8}
\begin{split}
&f_{12}(t,\v{k})= e^{-2t\sqrt{m(\v{k})^2 + t_1^2|\Omega(\v{k})|^2}}\\
&\cdot\left( \mbox{Tr}\left\{P_-(\v{k}) \de_{k_1}\hat{H}^0(\v{k}) P_+(\v{k}) \de_{k_2}\hat{H}^0(\v{k}) \right\}\right.\\
&\left.\ \ \ + \mbox{Tr}\left\{ P_-(\v{k})[A_1,\hat{H}^0(\v{k})]P_+(\v{k}) \de_{k_2}\hat{H}^0(\v{k}) \right\} \right),
\end{split}
\end{equation}
with $A_1= \Gamma_0(\v{k})\de_{k_1}\Gamma^\dg_0(\v{k})=\left(\begin{array}{cc}
0    & 0 \\
0   & i
\end{array}\right)$. 

We expand the following quantities around $\v{k}_F^-$,
\[
\begin{split}
& \hat{H}^0(\v{k}_F^- + \v{k}')-\mu= -\frac{3}{2}t_1\s_2 k_1' + \frac{3}{2}t_1 \s_1 k_2' + O(|\v{k}'|^2)\\
& P_-(\v{k}_F^-+\v{k}')= \frac{1}{2}\left(\begin{array}{cc}
 1    & -ie^{-i\text{arg}(\v{k}')} \\
i e^{i\text{arg}(\v{k}')}     & 1
\end{array}\right) + O(|\v{k}'|)\\
& P_+(\v{k}_F^-+\v{k}')= \frac{1}{2}\left(\begin{array}{cc}
 1    & ie^{-i\text{arg}(\v{k}')} \\
-ie^{i\text{arg}(\v{k}')}     & 1
\end{array}\right) + O(|\v{k}'|)\\
& \sqrt{m(\v{k}_F^-+\v{k}')^2 + t_1^2|\Omega(\v{k}_F^- + \v{k}')|^2}= \frac{3}{2} t_1|\v{k}'| + O(|\v{k}'|^2),
\end{split} 
\]
where $\s_1, \s_2,\s_3$ are the Pauli matrices and $\arg(\vec{k}')$ is the argument of $k_1'+ik_2'$. This allows us to expand \eqref{eq_8} and estimate
\begin{equation}
\label{eq_9}
\begin{split}
&\left|f_{12}(t,\v{k})+\frac{9}{8}t_1^2 \sin(2\text{arg}(\v{k}')) e^{-3tt_1|\v{k}'|}\right|\\
&\leq O\left(|\v{k}'| e^{-t t_1|\v{k}'|}\right) + O\left(t|\v{k}'|^2e^{-tt_1|\v{k}'|}\right),
\end{split}
\end{equation}
for $|\vec k'|$ sufficiently small. A similar computation for $f_{21}(t,\v{k})$ gives the same leading order, so that
\begin{equation}
\label{eq_10}
\begin{split}
\s_{12}^{(a)}&=O(\epsilon)-\lim_{p_0\rightarrow 0^-} \int_0^{\infty}dt \int_{B_{\e}(\v{0})} \frac{d\v{k}'}{(2\pi)^2} \\
&\cdot \left(\frac{e^{ip_0t}+ e^{-ip_0t}-2}{p_0}\right) \frac{9}{8}t_1^2\sin(2\text{arg}(\v{k}')) e^{-3tt_1|\v{k}'|},
\end{split}
\end{equation}
where the term bounded as $O(\epsilon)$ was obtained from the RHS of \eqref{eq_9}.
Using polar coordinates  $(|\v{k}'|,\text{arg}(\v{k}'))$, we see that the integral in \eqref{eq_10} vanishes by parity, and so $\s_{12}^{(a)}= O(\e)$.

\medskip

\textit{The contribution $\s_{12}^{(b)}$}. 
We closely follow Appendices A and B of \cite{GMP17}, which discuss the computation of $\s_{12}$ in the non-critical case. The reason these methods apply in the present case is that the energy bands do not touch on $\mc{B}\setminus B_{\e}$ (recall that $\s_{12}^{(b)}$ is the RHS of \eqref{eq_7} with the integration of $\vec{k}$ restricted to $\mc{B}\setminus B_{\e}$).

Note that for $\e>0$ fixed, $f_{12}(t,\v{k})$ and $f_{21}(t,\v{k})$ decay exponentially in $t$, uniformly in $\v{k}\in \mc{B}\setminus B_{\e}$. Hence, we can bring the limit inside the integral over $t$, 
\begin{equation}
\label{someeq203}
\s_{12}^{(b)}= -i\int_0^{\infty} dt\ t \int_{\mc{B}\setminus B_{\e}} \frac{d\v{k}}{(2\pi)^2} \left[ f_{12}(t,\v{k}) - f_{21}(t,\v{k}) \right].
\end{equation}
We also have that (cf.\ appendix A of \cite{GMP17}) 
\[
\begin{split}
&f_{ij}(t,\v{k})\\
&=\partial^2_t\mbox{Tr}\left\{ e^{t\tilde{H}^0(\v{k})} \tilde{P}_-(\v{k})\de_{k_i}\tilde{P}_-(\v{k}) e^{-t\tilde{H}^0(\v{k})}  \de_{k_j}\tilde{P}_-(\v{k}) \right\},
\end{split}
\]
where the trace decays exponentially to zero as $t\to\infty$, uniformly in $\vec k$. Using partial integration, this implies
\[
\int_0^{\infty} dt\ tf_{ij}(t,\v{k}) = \mbox{Tr}\left\{ \tilde{P}_-(\v{k})\de_{k_i}\tilde{P}_-(\v{k})  \de_{k_j}\tilde{P}_-(\v{k}) \right\}, 
\]
and using again the definition of $\Gamma_0(\vec k)$, we see that \eqref{someeq203} becomes
\begin{equation}
\label{eq_11}
\begin{split}
&\s_{12}^{(b)}= -i \int_{\mc{B}\setminus B_{\e}} \frac{d\v{k}}{(2\pi)^2}  \\
& \left( \mbox{Tr}\left\{ P_-(\v{k})[\de_{k_1}P_-(\v{k}), \de_{k_2} P_-(\v{k})]\right\} -\de_{k_2}\mbox{Tr}\left\{ A_1 P_-(\v{k})]\right\} \right),
\end{split}
\end{equation}
with $A_1$ as defined below \eqref{eq_8}.
We now write $P_-(\v{k})$ as $v_-(\v{k}) v^{\dg}_-(\v{k}) $, with
\[
v_-(\v{k})= \frac{1}{N(\v{k})}\left(\begin{array}{cc}
t_1\Omega^*(\v{k})  \\
\sqrt{m(\v{k})^2 + t_1^2|\Omega(\v{k})|^2} + m(\v{k}) 
\end{array}\right),
\]
and $N(\v{k})$ a normalization factor, equal to
\[
\left[ 2m(\v{k})^2 + 2t_1^2|\Omega(\v{k})|^2+ 2m(\v{k})\sqrt{m(\v{k})^2 + t_1^2|\Omega(\v{k})|^2} \right]^{\frac{1}{2}}. 
\]
It is easily checked that $v_-$ is real analytic over the whole Brillouin zone, except at the point $\v{k}_F^-$. Note that
\[
\mbox{Tr}\big\{P_-(\v{k}) [\de_{k_1}P_-(\v{k}), \de_{k_2}P_-(\v{k})] \big\}= \nabla\times\langle v_-(\v{k}), \nabla v_-(\v{k}) \rangle,
\]
which is proportional to the Berry curvature. 
By Stokes' theorem, \eqref{eq_11} becomes
\[
\begin{split}
&\s_{12}^{(b)}\\
&=\frac{i}{(2\pi)^2} \int_{\de B_{\e}}  d\v{k}\cdot \left[ \langle v_-(\v{k}), \nabla v_-(\v{k})\rangle + \mbox{Tr}\big\{A_1P_-(\v{k})\big\}\left(\begin{smallmatrix}
1\\
0\\
\end{smallmatrix}\right)  \right]\\
&=\frac{i}{(2\pi)^2} \int_{\de B_{\e}}  d\v{k}\cdot  \langle v_-(\v{k}), \nabla v_-(\v{k})\rangle + O(\e),
\end{split}
\]
where the line integral over the boundary of $B_\epsilon$ is run counterclockwise, and we used that $A_1P_-(\v{k})$ is uniformly bounded in matrix norm for the second step. Setting $\v{k}-\v{k}_F^-= \e(\cos\vth, \sin\vth)$, we can expand
\[
v_-(\v{k})= \frac{1}{\sqrt{2}}\left( \begin{array}{cc}
-ie^{-i\vth}  \\
1 
\end{array}\right) +O(\e).
\]
Combining the previous two equations, it follows that $
\s_{12}^{(b)}=1/4\pi+ O(\e)$, so that by letting \hbox{$\e\rightarrow 0$} we obtain the desired result
\[
\s_{12}= \s_{12}^{(a)}+\s_{12}^{(b)}= \frac{1}{4\pi}.
\]

\subsection{The general, interacting, case}
\label{section_univ}

The computation of $\s_{12}$ on the critical line, presented in the previous section, is 
based on an explicit evaluation of the current-current response function, which
cannot straightforwardly be generalized to the interacting model. In that case, correlation functions cannot be written in closed form; rather, they can be expressed in the form of infinite, yet convergent, series expansions, as reviewed below. In order to compute $\sigma_{12}$ on the dressed critical line, we use the fact that $\sigma_{12}$ is quantized and universal over each of the four disconnected regions of the interacting phase diagram shown in Fig.\ref{fig2}, as proved in \cite{GMP20}. As anticipated above, our goal is to show that the difference between the critical transverse conductivity and the symmetric average of the off-critical one, at a distance $\epsilon$ inside or outside the critical curve, vanishes as $\epsilon\to 0$, which implies the new part of our main result. 

\medskip

Let us describe our strategy more precisely. As before, without loss of generality, we focus on the 
critical line defined by $m^R_-(U,W,\phi)=0$ with $\phi\in(0,\pi)$. As in \cite[Sect.2.3.1]{GMP20}, we introduce the \textit{reference Hamiltonian}, 
defined in terms of $U,m_-^R,\phi$ as follows: 
\begin{equation}
\label{refhamiltonian}
\begin{split}
&\mc{H}^R_L(U,m^R_-,\phi):=\\
&  \mc{H}_{0,L}\left(m^R_-+3\sqrt{3}t_2\sin\phi, \phi\right) -3t_2\cos\phi\ \mc{N}_L + U\mc{V}_L \\
&\ \ \ +\xi(U,m^R_-,\phi)\sum_{\v{x}\in\Lambda_L}\psi^{\dg}_{\v{x}}\psi_{\v{x}}+ \delta(U,m^R_-,\phi)\sum_{\v{x}\in\Lambda_L}\psi^{\dg}_{\v{x}}\sigma_3\psi_{\v{x}}.
\end{split}
\end{equation}
where $\xi$ and $\delta$ are the chemical potential and mass counterterms, respectively, and $\sigma_3$ is the third Pauli matrix. The functions $\xi$ and $\delta$ are analytic in $U$, and should be thought of, respectively, as the shifts in $\mu$ and $W$ caused by the interaction (and thus vanishing for $U=0$). They are fixed so that the RG flow of the theory approaches a Gaussian fixed point in the infrared, characterized by a renormalized mass equal to $m^R_-$ (cf. \cite[Lemma 4.2]{GMP20}), and by a semimetallic behavior at $m^R_-=0$. Note that $\mc{H}^R_L(U,m^R_-,\phi)$ is a perturbation of the non-interacting Hamiltonian $\mc{H}_{0,L}\left(m^R_-+3\sqrt{3}t_2\sin\phi, \phi\right) -3t_2\cos\phi\ \mc{N}_L$, which is critical on the `correct', interacting
critical line $m^R_-=0$. We require that the reference Hamiltonian coincides
with $\mathcal H_{L}$ at parameters $W,\phi,\mu,U$, so that 
\begin{equation}
\label{someeq324}
\begin{split}
W&= m^R_-+3\sqrt{3}t_2\sin\phi + \delta(U,m^R_-,\phi)\\
\mu&= 3t_2\cos\phi - \xi(U,m^R_-,\phi).
\end{split}
\end{equation}
The function $m^R_-(U,W,\phi)$ that appears in Theorem \ref{thm_main} is nothing but the solution of the equation given by the first of \eqref{someeq324}, and the functions $\mf{z}(U,W,\phi), \mf{d}(U,W,\phi)$ correspond to $\xi(U,m^R_-,\phi)$ and $\delta(U,m^R_-,\phi)$ respectively, computed at $m^R_-=m^R_-(U,W,\phi)$. In other words, the reference Hamiltonian 
is just a rewriting of the original interacting Hamiltonian, re-expressed in terms of new parameters, in particular of the dressed, renormalized mass, which is a physically more natural parameter to deal with, particularly if we intend to study the system at, or close to, the dressed critical line. 

\medskip 

The multiscale construction of the counterterms and of the dressed critical 
line, via the inversion of \eqref{someeq324}, has been discussed in detail in 
\cite{GMP20}, and will be reviewed below, for the purpose of deriving the new 
results required for the proof of the main theorem in the critical case. 

By construction, the critical transverse conductivity we are interested in 
is equivalent to the transverse conductivity of the reference model at $m^R_-=0$, which we denote by 
\begin{equation}\label{critcond}\sigma^R_{12}(U,m^R_-,\phi)\big|_{m^R_-=0}\equiv \sigma^R_{12}(U,0,\phi).\end{equation} 
Our goal is to show that this quantity is equal to $1/4\pi$, exactly as in the non-interacting case. The key ingredient used to prove this is a representation of the transverse conductivity of the reference model at, or close to, the massless case, following from the multiscale expansion reviewed below and summarized in the following proposition. 

\begin{proposition}
\label{proposition_main}
There exists $U_0>0$ such that for $|U|\le U_0$, $\phi\in(0,\pi)$, 
and any sufficiently small, positive, $\epsilon$, 
the transverse conductivity of the reference model
$\s_{12}^R(U,m^R_-,\phi)$ with $-\epsilon\le m^R_-\le \epsilon$ can be decomposed as follows,
\begin{equation}
\label{someeq32}
\s_{12}^R(U,m^R_-,\phi)= \s_{12}^{R,(1)}(U,m^R_-,\phi)+\s_{12}^{R,(2)}(U,m^R_-,\phi).
\end{equation}
Here, the first term (the `singular part') is  
\begin{eqnarray}
&&\s_{12}^{R,(1)}(U,m,\phi)=\gamma^2\lim_{p_0\to 0^-}\frac1{p_0}\int_{\mathcal A_\epsilon}\frac{d\mb{k}}{(2\pi)^3} \cdot\label{eq_15}
\\
&&\qquad\qquad \cdot\textup{Tr}\left\{ \sigma_2 S_m(\mb{k})\sigma_1 \big({S}_{m}(\mb{k}+\mb{p})-S_m(\mb{k})\big) \right\},\nonumber
\end{eqnarray}
where $\gamma$ is an analytic function of $U$, equal to $\tfrac32t_1$ at $U=0$; $\mc{A}_\epsilon$ is the 3D ball of radius $\sqrt{\epsilon}$ centered at the origin; $\sigma_1$ and $\sigma_2$ are the first two Pauli matrices; $\mb{p}=(p_0,0,0)$, and 
the `dressed propagator' $S_m(\mb{k})$ at $\mb{k}=(k_0,k_1,k_2)$ is
\begin{equation} 
\label{eq_16}
{S}_{m}(\mb{k})= \left(\begin{array}{cc}
-iZ_{1} k_0 + m& v(ik_1+ k_2)\\
v(-ik_1+ k_2) & -iZ_{2}k_0 -m\\
\end{array}\right)^{-1},
\end{equation}
with $Z_1,Z_2,v$ three `dressed parameters' that are analytic functions of $U$, equal to $1$, $1$, $\frac{3}{2}t_1$ at $U=0$, respectively. 

Moreover, the second term in \eqref{someeq32} (the `regular part') is continuous at zero, in the sense that 
\begin{equation}
\label{eq_17}
\begin{split}
\lim_{\epsilon \rightarrow0^+}\s_{12}^{R,(2)}(U,\pm\epsilon,\phi)=
\s_{12}^{R,(2)}(U,0,\phi).
\end{split}
\end{equation}
\end{proposition}

The `singular' contribution $\sigma_{12}^{R,(1)}$ is nothing but 
the dressed lowest order diagram contribution in renormalized perturbation theory, which is the dominant contribution at low momenta. The dressing consists in a finite (analytic!) renormalization of the parameters $Z_1,Z_2,v$ 
in the interacting propagator, see \eqref{eq_16}, and of the vertex functions in directions $1$ and $2$, equal to $-\gamma\sigma_2$ and $\gamma\sigma_1$ respectively. In \eqref{eq_15}, $\mb{k}$ represents the quasi-momentum relative to the Fermi point $\vec k^-_F$; the restriction of $\mb{k}$ to $\mathcal A_\epsilon$ corresponds to the fact that the singular part only 
includes contributions from the infrared modes, at low quasi-momenta. The regular part of the conductivity includes all the contributions at larger quasi-momenta, as well as the (convergent!) sum of all the Feynman diagrams in renormalized perturbation theory involving a quartic electron-electron interaction; since the quartic interaction is irrelevant in the infrared (in an RG sense), all such terms are dimensionally better behaved in the infrared, as compared to the dominant contribution leading to \eqref{eq_15}, and this 
is ultimately the reason why the regular part is continuous on the critical line, as per \eqref{eq_17}. 

\medskip

Proposition \ref{proposition_main} is a corollary of the multiscale construction of the model, and it will be proved in the next section. Before 
we discuss it, let us prove here that the proposition readily implies the desired result for the critical conductivity. In order to compute \eqref{critcond}, we rewrite: 
\begin{equation}\label{critcondrew}
\begin{split}
&\s_{12}^{R}(U,0,\phi)=\tfrac12\lim_{\epsilon\to 0^+}\big(\s_{12}^{R}(U,\epsilon,\phi)+\s_{12}^{R}(U,-\epsilon,\phi)\big)\\
&\quad +\lim_{\epsilon\to 0^+}\big[\sigma_{12}^R(U,0,\phi)-\tfrac12\s_{12}^{R}(U,\epsilon,\phi)-\tfrac12\s_{12}^{R}(U,-\epsilon,\phi)\big].\end{split}
\end{equation}
Thanks to the validity of Theorem \ref{thm_main} in the off-critical case, 
already proved in \cite{GMP20}, the first term on the RHS equals 
$\tfrac1{4\pi}$ for all $\phi\in(0,\pi)$. We want to show that the second line vanishes, thanks to Proposition \ref{proposition_main}. We rewrite the conductivity as in \eqref{someeq32} and then, using the continuity property \eqref{eq_17}, notice that the only possible contribution to the second line of \eqref{critcondrew} comes from $\sigma_{12}^{R,(1)}$. 
In order to compute the contribution from $\sigma_{12}^{R,(1)}$, we recall \eqref{eq_15}, use \eqref{eq_16}, and find that
\begin{equation}\label{trace}\begin{split}
&\frac1{p_0}\textup{Tr}\left\{ \sigma_2 S_m(\mb{k})\sigma_1 \big({S}_{m}(\mb{k}+\mb{p})-S_m(\mb{k})\big) \right\}=\\
& =\frac{4v^2[Z_1Z_2(2k_0+p_0)-im(Z_1-Z_2)]k_1k_2}{D_m^2(\mb{k})D_m(\mb{k}+\mb{p})}\\
&-\frac{m(Z_1+Z_2)}{D_m(\mb{k})D_m(\mb{k}+\mb{p})},
\end{split}\end{equation}
where $D_m(\mb{k})=Z_1Z_2k_0^2+v^2(k_1^2+k_2^2)+m^2-im(Z_1-Z_2)k_0$; next, we plug \eqref{trace} into \eqref{eq_15}, thus getting, for $m=0$ and $m=\pm\epsilon$ respectively: 
\[ \sigma_{12}^{R,(1)}(U,0,\phi)=0\]
and 
\[ \sigma_{12}^{R,(1)}(U,\pm\epsilon,\phi)=\mp\epsilon(Z_1+Z_2)
\int_{\mathcal A_\epsilon}\frac{d\mb{k}}{(2\pi)^3}\frac1{\big[D_{\pm\epsilon}(\mb{k})\big]^2}.
\]
The value of the integral in the last equation is independent
 of the choice of the sign in the subscript of $D_{\pm\epsilon}(\mb{k})$. 
 Therefore, $\sigma_{12}^{R,(1)}(U,\epsilon,\phi)+\sigma_{12}^{R,(1)}(U,-\epsilon,\phi)=0$, which implies that the second line of 
\eqref{critcondrew} vanishes as announced. 

\begin{observation}\label{remark:1}
Note that the fact that $\sigma_{12}^{R,(1)}(U,0,\phi)=\frac12\big(\sigma_{12}^{R,(1)}(U,\epsilon,\phi)+\sigma_{12}^{R,(1)}(U,-\epsilon,\phi)\big)=0$ ultimately 
follows from an emergent parity symmetry of the effective relativistic infrared theory, which implies that $\sigma_{12}^{R,(1)}(U,\epsilon,\phi)$ is odd in $\epsilon$, thus enforcing 
that the right average to use in \eqref{critcondrew} is the standard arithmetic one, rather than other, more general, weighted averages.\end{observation}
 
Putting things together, we proved that 
\[\sigma_{12}^R(U,0,\phi)=\frac1{4\pi},
\]
as desired. We are left with discussing the proof Proposition \ref{proposition_main}, 
which is described in the next section. 

\section{Proof of Prop.\ \ref{proposition_main}}
\label{section_proof}
In this section we prove Proposition \ref{proposition_main}, whose result, as 
already mentioned, is a corollary of the multiscale RG construction of the 
interacting ground state, already discussed in detail in 
\cite{GJMP16,GMP20}. In order to explain the origin of the decomposition in 
\eqref{someeq32} and of the continuity of $\sigma_{12}^{R,(2)}$ at $m^R_-=0$, 
we first review the main aspects of the RG construction, at a level of detail sufficient for us to state and justify the main computations and estimates behind the proof of Proposition \ref{proposition_main}, referring the reader to \cite{GJMP16,GMP20} for additional details. Next, we use the output of the RG construction to prove the main content of the 
proposition of interest, that is the continuity bound \eqref{eq_17}. We assume, as before, that $\phi\in (0,\pi)$, and we take $|m_-^R|$ sufficiently small, so that 
$m_+^R:=m_-^R+6\sqrt3 t_2\sin\phi>|m_-^R|$. We stress that the method we use not only allows us to account for the effect of interactions, that is, it is robust under smooth modifications of the quartic part of the Hamiltonian (i.e., the interaction), but it is also flexible enough to deal with smooth modifications of the quadratic part of the Hamiltonian (i.e., the hopping): as clear from the discussion below, the only thing we really use is the structure of the relevant and marginal terms, see, e.g., eq.\eqref{absorbing} below, which ultimately follows from the symmetry properties of the model and the number of (quasi-)massless degrees of freedom at or close to the transition line. Therefore, even if the proof is spelled out only in the case of the interacting Haldane model, we expect it can be generalized to a larger class of interacting electron systems with critical semimetallic state; this will be the subject of a future publication.

\subsection{Review of the RG construction}

The starting point is a reformulation of the reference model \eqref{refhamiltonian} in terms of an interacting Grassmann integral. The generating functional $\mc{W}$ of multipoint field and current correlations 
at inverse temperature $\beta$ and finite volume can be written as
\begin{equation}
\label{gen_funct}
\mc{W}(\vphi, A):= \log \int P(d\psi) e^{- V(\psi)+ (\psi,\vphi) + (j,A) },
\end{equation}
where $\psi^{\pm}_{\mb{x},\r}$ with $\mb{x}\in [0,\beta)\times \L_L$ and $\r\in\{1,2\}$ are Grassmann variables, $\vphi$ is a Grassmann field conjugated to $\psi$, and $A$ is a real field conjugated to the current $j$, to be defined momentarily. In \eqref{gen_funct}, $P(d\psi)$ is the Gaussian Grassmann integration with propagator
\begin{equation}
g(\mb{x}-\mb{y})= \int \frac{d\mb{k}}{2\pi |\mc{B}|} e^{-i\mb{k}\cdot(\mb{x}-\mb{y})}\hat{S}^0(\mb{k}), 
\label{eq:gpropcompl}\end{equation}
where $\mb{x}=(x_0,\vec x)$, $\mb{k}=(k_0,\vec{k})$, 
$\hat S^0(\mb{k})$ was defined in \eqref{eq_1} and $\int \frac{d\mb{k}}{2\pi |\mc{B}|}$ is shorthand for the Riemann sum 
$\tfrac{1}{\beta L^2}\sum_{\mb{k}}$ with the sum over $\mb{k}$ running over 
$\tfrac{2\pi}\beta\mathbb Z\times \mathcal B_L$, see \eqref{mBL}. 
Moreover, 
\begin{equation}\label{interaction}
\begin{split}
V(\psi)&=\int_0^\beta dx_0\Big[ U\sum_{\substack{\v{x},\v{y}\in\L_L \\ \r,\r'=1,2}} n_{(x_0,\v{x}),\r}v_{\r,\r'}(\v{x}-\v{y}) n_{(x_0,\v{y}),\r'}\\
& + \sum_{\v{x}\in\L_L}\big( \xi(n_{\mb{x},1}+n_{\mb{x},2}) + \delta(n_{\mb{x},1}-n_{\mb{x},2})\big)\Big],
\end{split}
\end{equation}
with $n_{\mb{x},\r}= \psi^+_{\mb{x},\r}\psi^-_{\mb{x},\r}$, and $\xi=\xi(U,m^R_-,\phi)$, $\delta=\delta(U,m^R_-,\phi)$ are the same as in \eqref{refhamiltonian}. Finally, 
\begin{equation} \label{scalarproducts}
\begin{split}
(\psi,\vphi)&= \int_0^\beta dx_0\sum_{\v{x}\in\L_L}\left(\psi^+_{\mb{x}}\vphi^-_{\mb{x}}+ \vphi^+_{\mb{x}}\psi^-_{\mb{x}}\right) \\
(j,A)&= \sum_{\mu=0,1,2}\int\slashed{d}\mb{p} \, \hat{\jmath}_{\mb{p},\mu}\hat{A}_{\mb{p},\mu} , \end{split}
\end{equation}
where $\int \slashed{d}\mb{p}$ is shorthand for the Riemann sum 
$\tfrac{1}{\beta L^2}\sum_{\mb{p}}$ with the sum over $\mb{p}$ running over 
$\tfrac{2\pi}\beta\mathbb Z\times \mathcal D_L$, see the line after \eqref{currentoperator}; moreover, 
$\hat{\jmath}_{\mb{p},\mu}= \int \frac{d\mb{k}}{2\pi|\mc{B}|} \hpsi^+_{\mb{k}+\mb{p}}\Gamma_{\mu}(\v{k},\v{p})\hpsi^-_{\mb{k}}$, with the bare vertex functions $\Gamma_{\mu}$ defined as in \eqref{eqGammas} and following lines. 

The generating functional can be calculated in the thermodynamic and zero-temperature limits using the constructive RG approach 
described in \cite[Sect.III.C]{GJMP16} and in \cite[Sect.4]{GMP20}. This 
allowed us to prove that the thermodynamic functions, such as the free energy per site, and the Euclidean correlation functions, also known as Schwinger functions, are 
{\it analytic} functions of $U$, uniformly in $\beta,L$, and that their limits as 
$L\to\infty$ first and then $\beta\to\infty$ are analytic in $U$ in a small complex neighborhood of the origin. Moreover, the method also allows us to compute the asymptotic behavior of correlations at large space-imaginary time distances. The fact that renormalized perturbation theory converges is an important fact, based on combinatorial and analytic techniques that take advantage of the fermionic nature of the model (it would not be true for 
a theory involving bosonic degrees of freedom): it allows us to fully keep track of the effects of the lattice and, more in general, of the irrelevant terms appearing in the original Hamiltonian or generated under the RG flow by the iterative, multiscale integration scheme, which may affect the values of finite quantities, such as the Kubo conductivity. 

The RG procedure is based on an iterative integration procedure of the infrared modes in \eqref{gen_funct}: we first rewrite 
the propagator $g({\bf x})$ in \eqref{eq:gpropcompl} with ${\bf x}=(x_0,\vec x)$ as $g({\bf x})=g^{(1)}({\bf x})+\sum_{\omega=\pm}e^{-i\vec k^-_F\cdot\vec x}g^{(\le 0)}_\omega({\bf x})$, 
see \cite[eq.(29)]{GJMP16}, where the Fourier transform of $g^{(1)}$ (resp.\ $g^{(\le 0)}_\omega$) is supported on the complement of two small balls centered at $(0,\vec k_F^+)$ and $(0,\vec k^-_F)$ (resp.\ on a small ball centered at $(0,\vec k^\omega_F)$). Note that the Fourier transform of $g^{(1)}$ does not have singularities in the Brillouin zone, so that  
$g^{(1)}$ and all its derivatives decay to zero faster than any power as $|{\bf x}|$ becomes large. 
Correspondingly, we decompose the Grassmann field $\psi$ in \eqref{gen_funct} as $\psi_{\bf x}^\pm=\psi^{\pm\,(1)}_{\bf x}+\sum_{\omega=\pm}e^{\pm i\vec k^\omega_F\cdot\vec x}\psi^{\pm\,(\le 0)}_{\omega,{\bf x}}$, where $\psi^{(1)}, \psi^{(\le 0)}_+, \psi^{(\le 0)}_-$ are independent Grassmann fields, with propagators $g^{(1)}, g^{(\le 0)}_+, g^{(\le 0)}_-$, and we integrate out $\psi^{(1)}$, thus re-expressing $\mathcal W(\varphi,A)$ in terms of a Grassmann functional integral involving integration only with respect to $\psi^{(\le 0)}_\pm$, a new effective potential $V^{(0)}(\psi^{(\le 0)})$ replacing $V(\psi)$ and a new source term $B^{(0)}(\varphi,A,\psi^{(\le 0)})$ replacing $(\psi,\varphi)+(j,A)$; see \cite[eq.(30)]{GJMP16} or 
\cite[eq.(4.6)]{GMP20}. 

Next, we start the integration of the infrared degrees of freedom, which is performed iteratively, by using the decomposition $\psi^{(\le 0)}_\omega$ as
$\psi^{\pm\,(\le 0)}_{\omega,{\bf x}}=\sum_{h<h'\le 0} \psi^{\pm\,(h')}_{\omega,{\bf x}}+\psi^{\pm\,(\le h)}_{\omega,{\bf x}}$, which, for $\omega=+$ (resp.\ $\omega=-$), 
is valid for any $h\ge h^{*}_1$ (resp.\ $h\ge h^{*}_2$), with $h^{*}_1=\lfloor \log_2 m^R_+\rfloor$, with $m^R_+=m^R_-+6\sqrt3 t_2\sin \phi$, see \cite[eq.(4.24)]{GMP20} (resp.
$h^*_2=h^*_2(m_-^R):=\lfloor \log_2 |m^R_-|\rfloor$, to be interpreted as being $-\infty$ if $m^R_-=0$). The single-scale Grassmann field $\psi^{(h')}_\omega$ has propagator that, in momentum space, is supported on momenta ${\bf k}'$ such that $|{\bf k}'|$ is of the order $2^{h'}$, that is, it is bounded from above and below proportionally to $2^{h'}$; physically, ${\bf k}'$ represents the quasi-momentum, i.e., the crystalline momentum relative to the Fermi point ${\bf k}^-_F=(0,\vec k^-_F)$). We integrate out the 
fields on scales $0,-1,-2,\ldots$; once we get to scale $h^{*}_1$, we fully integrate out the field $\psi^{(\le h^{*}_1)}_+$, which is massive on that scale, see \cite[eq.(4.32) and the lines after]{GMP20}. We then iteratively continue the integration of the fields $\psi^{(h)}_-$, with $h^*_2<h\le h^*_1$; if $m^R_-\neq0$, once we get to scale $h^*_2$, we fully integrate out the field
$\psi^{(\le h^*_2)}_2$; if $m_R^-=0$, then $h^*_2=-\infty$ and the iteration continues for infinitely many steps. 

After the integration of the higher-scale degrees of freedom, say for definiteness of the scales higher than $h$, with $h^*_2<h\le h^*_1$, 
the generating function takes the form: 
\begin{eqnarray}  \mathcal W(\varphi,A)&=&\sum_{h'>h} S^{(h')}(\varphi,A)\label{equazione35}\\
&+&\log \int P(d\psi^{(\le h)}_-)e^{-V^{(h)}(\psi^{(\le h)}_-)+B^{(h)}(\varphi, A, \psi^{(\le h)}_-)},\nonumber
\end{eqnarray}
where $S^{(h')}(\varphi,A)$ is the single-scale contribution to the generating function, while the effective potential $V^{(h)}$
and the effective source term $B^{(h)}$ explicitly depend on the fluctuation field $\psi^{(\le h)}_-$, i.e., $V^{(h)}(0)=B^{(h)}(\varphi, A,0)=0$ (the source term also explicitly  depends on the external fields: $B^{(h)}(0,0,\psi^{(\le h)}_-)=0$). $P(d\psi^{(\le h)}_-)$ is the Grassmann Gaussian integration with propagator that, in Fourier space, reads 
\begin{eqnarray}
\label{eq34} &&
\hat g^{(\le h)}_-({\bf k}')=\chi_h({\bf k}')\\
&&\ \ \cdot \begin{pmatrix} -iZ_{1,h}k_0'+m^R_- & {-\frac{2}{3}v_h\Omega^*(\vec{k}_F^- +\vec{k}')} \\{-{\frac{2}{3}v_h\Omega(\vec{k}_F^- +\vec{k}')}} & -iZ_{2,h}k_0'-m^R_- \end{pmatrix}^{-1}  (1+O({\bf k}')),\nonumber
\end{eqnarray}
where: $\chi_h({\bf k}')\equiv \chi_0(2^{-h}{\bf k}')$ is a smooth cutoff function supported on quasi-momenta $|{\bf k}'|<2^{h}$ ($\chi_0$ is a smooth version of the characteristic function of a small ball centered at the origin; we choose it to be a compactly supported $C^\infty$-function, supported on the ball of radius $1/3$ centered at the origin, and identically equal to $1$ on the 
ball of radius $1/4$ centered at the origin); $Z_{1,h}, Z_{2,h}$ are the effective wave function renormalizations, and $v_h$ is the effective Fermi velocity, which converge exponentially fast to their infrared limits, i.e., there exist $Z_1(m^R_-),Z_2(m^R_-),v(m^R_-)$, analytic in $U$, such that $|Z_{i,h}-Z_i(m^R_-)|, |v_h-v(m^R_-)|\le C_\theta|U|2^{\theta h}$, for any $0\le \theta<1$ and some $C_\theta>0$, diverging as $\theta\to 1^-$; the dependence upon $m^R_-$ of these effective constants is continuous: $|Z_i(m^R_-)-Z_i(0)|\le C_\theta|U||m^R_-|^\theta$ and $|v(m^R_-)-v(0)|\le C_\theta|U||m^R_-|^\theta$, for any $0< \theta<1$ and some $C_\theta>0$. 

\begin{observation}The constants $Z_1,Z_2,v$ in the statement of Proposition \ref{proposition_main} are nothing but the values of the corresponding constants at zero mass: 
$Z_i:=Z_i(0)$ and $v:=v(0)$. Similarly, $\gamma:=\gamma(0)$, where $\gamma(m^R_-)$ is the mass-dependent effective vertex function defined below, see the discussion after eq.\eqref{ebruciavan}.\label{obs:m=0}\end{observation}

{The function $\Omega$ that appears in the propagator \eqref{eq34} is the same as that in the Bloch Hamiltonian \eqref{eq33}. For $|\vec{k}'|\ll1$, one has $\Omega(\vec{k}_F^{\omega}+\vec{k}')= \frac{3}{2}(ik_1'+\omega k_2') + O(|\vec{k}'|^2)$, so in principle in the RHS of \eqref{eq34} we could replace $\Omega$ by its linearization, and include the higher-order terms in the remainder $O(|\bf k'|)$. However, we prefer to retain the full lattice function $\Omega(\vec{k}_F^{\omega}+\vec{k}')$ in order to explicitly preserve the discrete rotational symmetry of \cite[eq.(4.7)]{GMP20}, which would be violated if $\Omega$ was replaced by its linear approximation. 

The effective potential $V^{(h)}$ in \eqref{equazione35} can be represented as follows: 
\begin{eqnarray} \hskip-.1truecm && V^{(h)}(\psi)=\!\int\! \tfrac{d{\bf k}'}{(2\pi)^3}\!\big[2^h\xi_{-,h}\hat\psi^+_{{\bf k}'}\hat\psi^-_{{\bf k}'}+2^h\delta_{-,h}\hat\psi^+_{{\bf k}'}\sigma_3\hat\psi^-_{{\bf k}'}\nonumber\\
&&\ +\hat\psi^+_{{\bf k}'}\begin{pmatrix} -iz_{1,-,h}k_0' & {-{\frac{2}{3}u_{-,h}\Omega^*(\vec{k}_F^- +\vec{k}')}} \nonumber\\
{-{\frac{2}{3}u_{-,h}\Omega(\vec{k}_F^- +\vec{k}')}} & -iz_{2,-,h}k_0' \end{pmatrix}\hat\psi^-_{{\bf k}'}\big]\\ 
&&\ +{\mathcal R} V^{(h)}(\psi)\label{absorbing},\end{eqnarray}
where: $\xi_{-,h}, \delta_{-,h}$ are two running counterterms, satisfying $|\xi_{-,h}|,\, |\delta_{-,h}|\le C_\theta|U|2^{\theta h}$ for $0\le \theta<1$ and $C_\theta>0$;
$z_{1,-,h}, z_{2,-,h}, u_{-,h}$ are the single-step contributions to the effective couplings $Z_{1,-,h}, Z_{2,-,h}, v_h$, satisfying the same bounds as the running counterterms, i.e., 
$|z_{1,-,h}|, |z_{2,-,h}|, |u_{-,h}|\le C_\theta|U|2^{\theta h}$; $\mathcal R V^{(h)}$ is the irrelevant part of $V^{(h)}$, which is an infinite linear combination of even monomials 
of the form $\int d{\bf x}_1\cdots d{\bf x}_{n} W_{n,p}^{(h)}(\underline{\bf x};\underline{\alpha}) D^{\underline{\alpha}}\Psi(\underline{\bf x})$, 
where ${\bf x}_i=(x_{0,i},\vec x_i)$, $\int d{\bf x}_i$ is shorthand for $\int_0^\beta x_{0,i}\sum_{\vec x_i\in\Lambda}$, $\underline\alpha=((p_1,i_1),\ldots,(p_n,i_n))$, 
$\underline i=(i_1,\ldots,i_n)$, $\underline{\bf x}=({\bf x},\ldots,{\bf x}_n)$, and $D^{\underline{\alpha}}\Psi(\underline{\bf x})=
\partial^{p_1}_{i_1}\psi^+_{{\bf x}_1}\partial^{p_2}_{i_2}\psi^-_{{\bf x}_2}\cdots 
\partial^{p_{n-1}}_{i_{n-1}}\psi^+_{{\bf x}_{n-1}}\partial^{p_n}_{i_n}\psi^-_{{\bf x}_n}$ with $p_j=0,1$ and $i_j=0,1,2$: irrelevance of the monomial means that $n$ and 
$p:=\sum_{j=1}^np_j$ are such that the {\it scaling dimension} of the monomial
$$d_{sc}(n,p)=3-n-p$$
is negative; finally, the kernels of the irrelevant terms satisfy $\| W^{(h)}_{n,p}\|_h\le C^n|U|^{{\max}\{1,\frac{n}2-1\}}2^{d_{sc}(n,p)h}$, where 
$\| W^{(h)}_{n,p}\|_h:=\int d{\bf x}_2\cdots d{\bf x}_n |W^{(h)}_{n,p}(\underline{\alpha};\underline{\bf x})| e^{\kappa\sqrt{2^h\delta(\underline{\bf x})}}$ with $\kappa$ a sufficiently small positive constant and $\delta(\underline{\bf x})$ the tree distance among the elements of $\underline{\bf x}$ (also known as the Steiner diameter of $\{{\bf x}_1,\ldots,{\bf x}_n\}$, see 
\cite[footnote 19]{GMR21}). 

Similarly, the effective source term $B^{(h)}$ can be represented as: 
\begin{equation} B^{(h)}(\varphi,A,\psi)=(\psi,\varphi)+(J_h,A)+\mathcal R B^{(h)}(\varphi,A,\psi),\label{ebruciavan}\end{equation}
where, using a notation similar to \eqref{scalarproducts} and the lines after, 
 $(J_h,A)=\sum_{\mu=0,1,2}\int\slashed{d}\mb{p} \, \hat{J}_{h,\mu}(\mb{p})\hat{A}_{\mb{p},\mu}$; here $\hat{J}_{h,\mu}(\mb{p})$ is the dressed current, of the form 
$\hat{J}_{h,\mu}(\mb{p})= \int \frac{d\mb{k}'}{2\pi|\mc{B}|} \hpsi^+_{\mb{k}'+\mb{p}}\Gamma_{h,\mu}(\vec{k}',\vec{p})\hpsi^-_{\mb{k}'}$, with $\Gamma_{h,\mu}(\vec{k},\vec{p})$ 
the dressed vertex functions, such that $\Gamma_{h,0}(\vec{0},\vec{0})=\begin{pmatrix}\zeta_{1,h} & 0 \\ 0 & \zeta_{2,h}\end{pmatrix}$, $\Gamma_{h,1}(\vec{0},\vec{0})=-\gamma_h \sigma_2$, $\Gamma_{h,2}(\vec{0},\vec{0})=\gamma_h \sigma_1$, $|\Gamma_{h,\mu}(\vec{k},\vec{p})-\Gamma_{h,\mu}(\vec{0},\vec{0})|\le C(|\vec k|+|\vec p|)$ for some $C>0$, 
and $\partial_{m_-^R}\Gamma_{h,\mu}(\vec{k},\vec{p})=\partial_{m_-^R}\Gamma_{h,\mu}(\vec{0},\vec{0}) =O(U 2^{(\th-1)h})$;
the effective vertex constants $\zeta_{1,h}$, $\zeta_{2,h}$, 
$\gamma_h$ are analytic functions of $U$, uniformly in $h$, equal to $1,1,\tfrac32t_1$ at $U=0$ respectively; as $h\to-\infty$, the effective vertex constants converge to their infrared limits, denoted by $\zeta_1(m^R_-),\zeta_2(m^R_-),\gamma(m^R_-)$, which are continuous in $m^R_-$: 
$|\zeta_i(m^R_-)-\zeta_i(0)|\le C_\theta|U||m^R_-|^\theta$ and $|\gamma(m^R_-)-\gamma(0)|\le C_\theta|U||m^R_-|^\theta$, for any $0< \theta<1$ and some $C_\theta>0$
(as stated in Remark \ref{obs:m=0}, we let $\gamma:=\gamma(0)$ and, similarly, $\zeta_i:=\zeta_i(0)$); by a Ward identity (see \cite[eq.(20)]{GJMP16}, one has $\zeta_i=-Z_i$ and $\gamma=v$; the speed of convergence to the limit is exponential, i.e., 
$|\zeta_{i,h}-\zeta_i(m^R_-)|, |\gamma_h-\gamma(m^R_-)|\le C_{\theta}|U|2^{\theta h}$ for any $0<\theta<1$ and some $C_\theta>0$; moreover, $\mathcal R B^{(h)}$ is the irrelevant part of $B^{(h)}$, 
which is an infinite linear combination of irrelevant monomials of the form 
$\int d\underline{\bf y}\, d\underline{\bf z}\, d \underline{\bf x}W_{l,m,n,p}^{(h)}(\underline{\bf y}, \underline{\bf z}, \underline{\bf x};\underline\mu,\underline{\alpha}) \varphi(\underline{\bf y})A_{\underline\mu}(\underline{\bf z})D^{\underline{\alpha}}\Psi(\underline{\bf x})$, where we use a notation analogous to the one used above for $\mathcal R V^{(h)}$;
irrelevance of the monomial means that its scaling dimension is negative, where, in the presence of external fields, the scaling dimension of a monomial of order $l$ in $\varphi$, $m$ in $A$, $n$ in $\psi$, with $p$ derivatives acting on the $\psi$ fields, should be modified into $$d_{sc}(l,m,n,p):=3-2l-m-n-p;$$ finally, the kernels $W_{l,m,n,p}^{(h)}$ satisfy a norm bound analogous to that of the kernels of the effective potential, namely \begin{equation}\|W_{l,m,n,p}^{(h)}\|_h\le 
C^{l+m+n}2^{d_{sc}(l,m,n,p)h}.\label{normboundB}\end{equation}

In \eqref{equazione35}, the single-scale contribution to the generating function, $W^{(h)}$, admits a representation similar to $V^{(h)}$ and $B^{(h)}$, that is, it can be written as 
an infinite linear combination of monomials of the form $\int d\underline{\bf y}\, d\underline{\bf z}\,W_{l,m,0,0}^{(h)}(\underline{\bf y}, \underline{\bf z};\underline\mu) \varphi(\underline{\bf y})A_{\underline\mu}(\underline{\bf z})$, where the kernels $W_{l,m,0,0}^{(h)}$ satisfy the same norm bound as \eqref{normboundB}, with $n=p=0$. 

At each step of the iteration leading to \eqref{equazione35}, the marginal quadratic terms in $\psi$, i.e., those in the second line of \eqref{absorbing}, are re-absorbed into the Gaussian Grassmann measure, thus
contributing to the iterative dressing of the propagator. Next, the modified propagator is decomposed into a single-scale contribution, associated with momenta on scale $h$, plus an 
infrared contribution, associated with momenta on scales smaller than $h$; the contribution from the single-scale propagator at scale $h$ is integrated out, and a formula analogous to 
\eqref{equazione35} is obtained, with $h$ replaced by $h-1$. 

The iterative construction sketched above induces a {\it convergent} expansion in $U$, uniformly in $m_-^R$ as ${m_-^R\to 0}$, for the kernels $W_{l,m,n,p}^{(h)}$ 
(for uniformity of notation, from now on we denote by $W_{0,0,n,p}^{(h)}$ the kernels of $V^{(h)}$, which were previously denoted by $W^{(h)}_{n,p}$). These can conveniently be expressed as a sum over the values of {\it Gallavotti--Nicol\`o} trees, identical to those in \cite[Section 3]{GM10}, modulo the minor differences described in
\cite{GMP20} in the paragraph right before \cite[Lemma 4.1]{GMP20}. We shall write
$$W_{l,m,n,p}^{(h)}=\sum_{\tau\in \mathcal T_{l,m,n,p}^{(h)}}W_{l,m,n,p}[\tau],$$
where $W_{l,m,n,p}[\tau]$ is the tree value of $\tau$. The tree expansion is a combinatorially convenient (and better behaved) resummation of the more naive
expansion in Feynman diagrams: each tree value is nothing but the sum over a family of Feynman diagrams with given vertices (represented by the endpoints of the Gallavotti--Nicol\`o trees)
and given scale labels associated with the propagators, satisfying a hierarchical structure of clusters into clusters compatible with the structure of the tree (a cluster is a Feynman sub-diagram with propagators whose scale labels are all higher than those of the lines exiting from the sub-diagram itself); see \cite[Section 5]{GeM01} for a review. 
In particular, each (rooted) tree $\tau\in\mathcal T_{l,m,n,p}^{(h)}$ contributing to 
$W^{(h)}_{l,m,n,p}$ has root on scale $h$ and endpoints on higher scales, between $h+1$ and $1$; the endpoints on scales lower than $1$ correspond to `dressed' vertices, associated with either one of the two contributions in the first line of \eqref{absorbing} or one of the first two contributions on the RHS of \eqref{ebruciavan};
the endpoints on scale $1$ correspond to `bare' vertices, associated with one of the contributions of the `bare' action $-V(\psi)+(\varphi,\psi)+(j,A)$ in the exponent 
on the RHS of \eqref{gen_funct}. Note that any endpoint associated with a quartic contribution (the first on the RHS of \eqref{interaction}) is necessarily on scale 1; this convention is related to the irrelevance of the quartic interaction: $d_{sc}(0,0,4,0)=-1$.  We shall refer to the endpoints associated with a quartic interaction or to one of the running counterterms (see \eqref{absorbing} and following lines) as `interaction endpoints'.
The following key estimate on the sum of the tree values over trees having at least one interaction endpoint is valid, and will be used in the rest of the proof: 
\begin{equation}\label{shortm}\sum_{\substack{\tau\in \mathcal T_{l,m,n,p}^{(h)}: \\ N_{\text{int}}(\tau)\ge 1}}\!\!\!\!\!\!\!\|W_{l,m,n,p}[\tau]\|_h\le C_{\theta}^{l+m+n}|U|2^{d_{sc}(l,m,n,p)h}2^{\theta h},\end{equation}
where $N_{\text{int}}(\tau)$ is the number of interaction endpoints, 
$0\le \theta<1$, and $C_\theta$ is a positive constant such that $C_\theta\to+\infty$ as $\theta\to1^-$. The good factor $2^{\theta h}$ physically represents a dimensional 
gain coming from the fact that the quartic interaction is irrelevant in the infrared, with scaling dimension $-1$ (the `$1$' in the condition $\theta<1$ is the opposite of the scaling dimension of the quartic terms). 

\subsection{The differentiable and singular contributions}

In view of \eqref{def_Kubo}, we need to compute the left derivative of the current-current response function $\hat K_{12}(p_0)$ at $p_0=0$. 
Making the dependence upon $m_-^R$ explicit, from the RG construction we obtain the rewriting:
\begin{equation} 
\label{eq_22b}
\hat{K}_{12}(p_0;m^R_-)= \sum_{h=h^*_2}^0 \hat W^{(h)}_{0,2,0,0}((p_0,\vec 0);(1,2)),
\end{equation}
where $\hat W^{(h)}_{0,2,0,0}(\mb{p};(\mu_1,\mu_2))$ is the Fourier transform of the kernel of $W^{(h)}(\varphi,A)$ with $l=0$ and $m=2$ defined after \eqref{normboundB}.
As reviewed above, $\hat W^{(h)}_{0,2,0,0}$ can be written as a sum over Gallavotti--Nicol\`o trees, which induces the following decomposition of the current-current response function:
\begin{equation}\label{decK}\hat{K}_{12}(p_0;m^R_-)=\hat{K}_{12}^{\text{diff}}(p_0;m^R_-)+\hat{K}_{12}^{\text{sing}}(p_0;m^R_-),\end{equation}
where $\hat{K}_{12}^{\text{diff}}(p_0;m^R_-)$ (resp.\ $\hat{K}_{12}^{\text{sing}}(p_0;m^R_-)$) is the sum over the scales $h$ from $h^*_2$ to $0$ of the sum over trees $\tau\in 
\mathcal T_{0,2,0,0}^{(h)}$ with at least one interaction endpoint (resp.\ with no interaction endpoints) of $\hat W_{0,2,0,0}[\tau]((p_0,0);m^R_-)$. 
As we shall shortly see, the first contribution on the RHS of \eqref{decK} is continuously differentiable with respect to $p_0$, while the second is 
`singular', i.e., its derivative with respect to $p_0$ is dimensionally logarithmically divergent as $p_0\to 0$. As explained below, the decomposition \eqref{decK} is the basis for the rewriting \eqref{someeq32}. 

Let us first focus on the first term on the RHS of \eqref{decK}: from \eqref{shortm}, we find that 
\begin{equation*}
|\hat{K}_{12}^{\text{diff}}(p_0;m^R_-)|\le \sum_{h=h^*_2}^0C_\theta^2|U|2^{h(\theta+1)},\end{equation*}
which is $O(|U|)$, uniformly in $h^*_2$ as $h^*_2\to-\infty$ or, equivalently, in $m_-^R$ as $m_-^R\to 0$. The proof of the tree bounds leading to \eqref{shortm} readily implies that 
the derivative with respect to $p_0$ of the contribution to $\hat{K}_{12}^{\text{diff}}(p_0;m^R_-)$ from trees with root on scale $h$ can be bounded `dimensionally', i.e., 
in the same way as the corresponding contributions to $\hat{K}_{12}^{\text{diff}}(p_0;m^R_-)$ times an additional factor (const.)$2^{-h}$, which bounds the effect of the derivative (note, in fact that the 
contributions from trees with root on scale $h$ involve propagators on scales $>h$, i.e., supported on quasi-momenta of the order $2^{h}$ or larger):
$$|\partial_{p_0}\hat{K}_{12}^{\text{diff}}(p_0;m^R_-)|\le \sum_{h=h^*_2}^0C_\theta^2|U|2^{h\theta},$$
which is also $O(|U|)$, uniformly in $m_-^R$ as $m_-^R\to 0$. Moreover, $\partial_{p_0}\hat{K}_{12}^{\text{diff}}(p_0;m^R_-)$ is H\"older continuous in $m_-^R$ at $m_-^R=0$, for any exponent smaller than 1: the same 
`dimensional bounds' mentioned above show that the contribution to $\partial_{p_0}\hat{K}_{12}^{\text{diff}}(p_0;m^R_-)-\partial_{p_0}\hat{K}_{12}^{\text{diff}}(p_0;0)$ 
from trees with root on scale $h\ge h^*_2$ can be bounded 
in the same way as the corresponding contributions to $\partial_{p_0}\hat{K}_{12}^{\text{diff}}(p_0;m^R_-)$ times an additional factor (const.)$|m_-^R|2^{-h}$, which bounds the effect of making a first-order Taylor expansion in $m_-^R$; on the other hand, $\partial_{p_0}\hat{K}_{12}^{\text{diff}}(p_0;m^R_-)$ has no contributions from scales $h<h^*_2$, while the contribution to $\partial_{p_0}\hat{K}_{12}^{\text{diff}}(p_0;0)$ from any scale $h<h^*_2$ is bounded by $C^2_\theta|U|2^{h\theta}$, so that
\begin{eqnarray}&& |\partial_{p_0}\hat{K}_{12}^{\text{diff}}(p_0;m^R_-)-\partial_{p_0}\hat{K}_{12}^{\text{diff}}(p_0;0)| \label{derp0diff}\\
&&\qquad  \le |m_-^R|\sum_{h=h^*_2}^0C_\theta^2|U|2^{h(\theta-1)} +
\sum_{h<h^*_2}C_\theta^2|U|2^{h\theta}\nonumber\\
&&\qquad =C'_\theta |U| |m_-^R|2^{h^*_2(\theta-1)}+C''_\theta|U|2^{h^*_2\theta} =C'''_\theta |U| |m_-^R|^\theta,\nonumber\end{eqnarray}
where in the last identity we used the definition of $h^*_2:=\lfloor \log_2|m_-^R|\rfloor$. 

Let us now consider the `singular' term $\hat{K}_{12}^{\text{sing}}(p_0;m^R_-)$, which is by definition the contribution to the RHS of \eqref{eq_22b} 
from trees with no interaction endpoints and two endpoints associated with source terms $(J_h,A)$ and $(J_{h'},A)$, with $h,h'\ge h^*_2$. This contribution is very explicit: it is just the sum over scales of the dressed `bubble diagram':
\begin{center}
 \begin{tikzpicture}[baseline=-2pt]
  \begin{feynman}
    \vertex (a1) ;
    \vertex[right=1cm of a1] (a2);
    \vertex[right=2cm of a2] (a3);
    \vertex[right=1cm of a3] (a4) ;

    \diagram* {

      (a1) -- [boson, very thick, style=gray, momentum= $\mb{p}$] (a2),
      (a2) -- [fermion, half left, looseness=1.5, edge label=$\mb{k}'+\mb{p}$] (a3),
      (a3) -- [fermion, half left, looseness=1.5, edge label'=$\mb{k}'$] (a2),
      (a3) -- [boson, very thick, style=gray, momentum=$\mb{p}$] (a4),

    };
    \draw (1,0) node[ctVertex,E] {};
    \draw (3,0) node[ctVertex,E] {};
    \draw (1.5,0) node{$1,h'$};
    \draw (2.5,0) node{$2,h$};
\end{feynman}
\end{tikzpicture}
\end{center}
with $\mb{p}=(p_0,\vec 0)$, whose explicit expression is
\begin{equation}
\label{value_bubble}
\begin{split}
&-\int \frac{d\mb{k}'}{2\pi|\mathcal B|}\mbox{Tr}\Big\{\Gamma_{h\vee  h',1}(\vec{k}',\vec{p})
\hat{g}^{(h)}_{m_-^R}(\mb{k}')\\ &\qquad\qquad\qquad  \cdot
\Gamma_{h\vee h',2}(\vec{k}'+\vec{p},-\mb{p})\hat{g}^{(h')}_{m_-^R}(\mb{k}'+\mb{p}) \Big\}, 
\end{split}
\end{equation}
where $h\vee h':=\max\{h,h'\}$, the functions $\Gamma_{h,\mu}(\vec{k}',\vec{p})$ are the dressed vertex functions whose properties are spelled out after \eqref{ebruciavan}, and, if $h>h^*_2(m^R_-)$,
$\hat{g}^{(h)}_{m_-^R}(\mb{k}')$ is the single-scale counterpart of \eqref{eq34}, satisfying 
\begin{eqnarray}
\label{eq34bis} &&
\hat g^{(h)}_{m_-^R}({\bf k}')=f_h({\bf k}')\\
&&\ \ \cdot \begin{pmatrix} -iZ_{1,h}k_0'+m^R_- & v_h(ik_1'+k_2') \\  v_h(-ik_1'+k_2') &  -iZ_{2,h}k_0'-m^R_- \end{pmatrix}^{-1}  (1+O({\bf k}')),\nonumber
\end{eqnarray}
where $f_h(\mb{k}'):=\chi_h(\mb{k}')-\chi_{h-1}(\mb{k}')$ is supported on momenta of order $2^h$, i.e., in the region $\tfrac{2^h}8\le |\mb{k}'|\le \tfrac{2^h}3$ (if $h=h^*_2(m^R_-)$ we shall identify $\hat{g}^{(h)}_{m_-^R}(\mb{k}')$ with $\hat{g}^{(\le h)}_{-}(\mb{k}')$ in \eqref{eq34}). Moreover, the effective constants 
$Z_{1,h},Z_{2,h},v_h$ satisfy the properties described in the paragraph after \eqref{eq34}. For later reference, we note that, on their supports (which are the same as the support of $f_h$ or of $\chi_h$, depending on whether $h>h^*_2(m^R_-)$ or $h=h^*_2(m^R_-)$), the derivatives of $g^{(h)}_{m_-^R}(\mb{k}')$ 
with respect $k_0$ and/or $m_-^R$ satisfy the following dimensional bounds: 
\begin{equation}\begin{split}
& \big|\partial_{k_0'}\hat g^{(h)}_{m_-^R}({\bf k'})\big|,\ \big|\partial_{m_-^R}\hat g^{(h)}_{m_-^R}({\bf k'})\big|
  \le C2^{-2h}, \\
& \big|\partial_{k_0'}\partial_{m_-^R}\hat g^{(h)}_{m_-^R}({\bf k}')\big|\le C2^{-3h}.\end{split}\label{boundsgh} \end{equation}
Let us go back to the decomposition \eqref{decK}: plugging it into \eqref{def_Kubo}, we find that the transverse conductivity of the reference model can be rewritten as
\begin{equation}\label{decsigmafin}\sigma_{12}^R(U,m_-^R,\phi)=\sigma_{12}^{\text{diff}}(U,m_-^R,\phi)+\sigma_{12}^{\text{sing}}(U,m_-^R,\phi),\end{equation}
where we defined 
\begin{equation}\begin{split}
\sigma_{12}^{\text{diff}}(U,m_-^R,\phi)&=\frac1{|\vec\ell_1\times \vec \ell_2|}\partial_{p_0}\hat K_{12}^{\text{diff}}(0;m_-^R),\\
\sigma_{12}^{\text{sing}}(U,m_-^R,\phi)&=\frac1{|\vec\ell_1\times \vec \ell_2|}\partial_{p_0}^-\hat K_{12}^{\text{sing}}(0;m_-^R).\end{split}\label{definitions}\end{equation}
In particular, from the explicit expression of $\hat K_{12}^{\text{sing}}$, we find that 
\begin{equation}\label{decomposto.0} \sigma_{12}^{\text{sing}}(U,m_-^R,\phi)=-\!\!\sum_{h,h'=h^*_2}^0\!\!\partial^-_{p_0}\int \frac{d\mb{k}'}{(2\pi)^3}\mc{G}_{h,h'}(\mb{k}',\mb{p};m^R_-),\end{equation}
with $h^*_2=h^*_2(m_-^R)$ and 
\begin{eqnarray}\mc{G}_{h,h'}(\mb{k}',\mb{p};m^R_-)&=&
\mbox{Tr}\Big\{\Gamma_{h\vee  h',1}(\vec{k}',\vec{p})\hat{g}^{(h)}_{m_-^R}(\mb{k}')\label{defmcGhh}\\
&\cdot&\Gamma_{h\vee h',2}(\vec{k}'+\vec{p},-\vec{p})\hat{g}^{(h')}_{m_-^R}(\mb{k}'+\mb{p}) \Big\}.\nonumber\end{eqnarray}

\subsection{Completion of the proof}

We now use the decomposition \eqref{decsigmafin} and the bounds derived above to prove the continuity property \eqref{eq_17}. By definition, 
\begin{eqnarray} 
\s_{12}^{R,(2)}(U,m^R_-,\phi)&=&\sigma_{12}^{\text{diff}}(U,m_-^R,\phi)\\ 
&+&\big[\sigma_{12}^{\text{sing}}(U,m_-^R,\phi)-
\s_{12}^{R,(1)}(U,m^R_-,\phi)\big],\nonumber\end{eqnarray}
with $\sigma^{R,(1)}_{12}$ as in \eqref{eq_15}. Using the definition of $\sigma_{12}^{\text{diff}}$ (see the first line of \eqref{definitions}), and the H\"older continuity property 
\eqref{derp0diff}, we immediately find that $\lim_{m_-^R\to 0}\sigma_{12}^{\text{diff}}(U,m_-^R,\phi)=\sigma_{12}^{\text{diff}}(U,0,\phi)$. We are then left with proving that, for $m=\pm\epsilon$, 
\begin{equation}\label{expbrack}\begin{split} & \lim_{\epsilon\to 0^+}\big[\sigma_{12}^{\text{sing}}(U,m,\phi)-
\s_{12}^{R,(1)}(U,m,\phi)\\
& \qquad\ \ -\sigma_{12}^{\text{sing}}(U,0,\phi)+\s_{12}^{R,(1)}(U,0,\phi)\big]=0.\end{split}\end{equation}
For $m=\pm\epsilon$, recalling that $\mathcal A_\epsilon$ is the ball of radius $\sqrt\epsilon$ centered at the origin, we rewrite $\s_{12}^{R,(1)}(U,m,\phi)$ in \eqref{eq_15} as
\begin{equation}\s_{12}^{R,(1)}(U,m,\phi)\!=-\partial^-_{p_0}\!\!\!\sum_{h,h'=h_{1,\epsilon}}^{h_{2,\epsilon}}\int\limits_{\mathcal A_\epsilon}\! \frac{d\mb{k}'}{(2\pi)^3}
\mc{F}_{h,h'}(\mb{k}',\mb{p};m)\Big|_{\mb{p}=\mb{0}},\label{decompost}\end{equation}
where $h_{1,\epsilon}\equiv h^*_2(\epsilon)=\lfloor \log_2\epsilon\rfloor$, $h_{2,\epsilon}=\lfloor \log_2(9\sqrt\epsilon)\rfloor$, $\mb{p}=(p_0,\vec0)$, 
and, if $h,h'>h^*_2(m^R_-)$,
\begin{equation}\begin{split} \mathcal F_{h,h'}(\mb{k}',\mb{p};m)&=-\gamma^2 f_{h}(\mb{k}')f_{h'}(\mb{k}'+\mb{p})\\
&\cdot \textup{Tr}\left\{ \sigma_2 S_m(\mb{k}')\sigma_1 {S}_{m}(\mb{k}'+\mb{p})\right\},\end{split}\label{Eq:57}\end{equation} 
with $\gamma$ and $S_m({\bf k})$ the same as in \eqref{eq_15}, and where $f_h$ was defined after \eqref{eq34bis} (if $h=h^*_2(m^R_-)$, then $f_h$ in \eqref{Eq:57} should be replaced by $\chi_h$, and similarly for $h'$). Note that the upper bound in the sum over $h,h'$ in \eqref{decompost} is due to the fact that, for $\mb{p}$ small enough, the support of $\mathcal F_{h,h'}(\mb{k}',\mb{p};m)$ 
intersects $\mathcal A_\epsilon$ only if $h,h'\le h_{2,\epsilon}$. Note also that $\s_{12}^{R,(1)}(U,0,\phi)$ admits a decomposition similar to  \eqref{decompost}, 
with $h_{1,\epsilon}$ replaced by $-\infty$ and $\mathcal F_{h,h'}(\mb{k}',\mb{p};m)$ replaced by $\mathcal F_{h,h'}(\mb{k}',\mb{p};0)$.

We now plug the decompositions \eqref{decomposto.0} and \eqref{decompost}, as well as the analogous one for $\s_{12}^{R,(1)}(U,0,\phi)$, 
 in the expression in brackets in \eqref{expbrack}, 
and rewrite it as the sum of three terms, $I_1$, $I_2$, $I_3$, 
which will separately be shown to tend to zero as $\epsilon\to 0^+$. These are defined as follows
(recall that $\mb{p}=(p_0,\vec0)$, that $m=\pm\epsilon$, and let $h_{3,\epsilon}=\lfloor \log_2\sqrt\epsilon\rfloor$): 
\begin{equation}\begin{split} &
\hskip-.3truecm I_1\!=\!\! \!\!\sum_{h,h'=h_{3,\epsilon}}^0\int\limits_{\mathcal A_\epsilon^c}\! \frac{d\mb{k}'}{(2\pi)^3}\partial_{p_0}\Big[
\mc{G}_{h,h'}(\mb{k}',\mb{p};0)-\mc{G}_{h,h'}(\mb{k}',\mb{p};m)\Big]\Big|_{\mb{p}=\mb{0}}\\
& 
\hskip-.3truecm I_2\!=\!\! \!\!\sum_{h,h'=h_{1,\epsilon}}^{h_{2,\epsilon}}\int\limits_{\mathcal A_\epsilon}\! \frac{d\mb{k}'}{(2\pi)^3}\partial_{p_0}\Big[\mc{F}_{h,h'}(\mb{k}',\mb{p};m)-\mc{G}_{h,h'}(\mb{k}',\mb{p};m)\Big]\Big|_{\mb{p}=\mb{0}}\\
& 
\hskip-.3truecm I_3\!=\partial^-_{p_0}\!\!\sum_{h,h'=-\infty}^{h_{2,\epsilon}}\int\limits_{\mathcal A_\epsilon}\! \frac{d\mb{k}'}{(2\pi)^3}\Big[\mc{G}_{h,h'}(\mb{k}',\mb{p};0)-\mc{F}_{h,h'}(\mb{k}',\mb{p};0)\Big]\Big|_{\mb{p}=\mb{0}}\end{split}
\end{equation}
In the definition of $I_1$, the lower bound on the sum over $h,h'$ is due to the fact that, for $\mb{p}$ small enough, 
the support of $\mathcal G_{h,h'}(\mb{k}',\mb{p};m)$ 
intersects the complement of $\mathcal A_\epsilon$ only if $h,h'\ge h_{3,\epsilon}$). Note that in the definitions of $I_1$ and $I_2$, we could move the left derivatives
inside the sums, because they run over finitely many terms; moreover, each term in each of the two sums is smooth in $\mb{p}$ and absolutely integrable 
and, therefore, 
in both cases we could change the left derivatives to full derivatives and move them inside the integral. Note also that, without loss of generality, the sums over $h,h'$ 
in the definitions of $I_1,I_2$ can be restricted to pairs of scale labels such that $|h-h'|\le 1$: in fact, if $|h-h'|\ge 2$ and $\mb{p}$ is small enough, the summands are zero, due to the compact support properties of 
$\mathcal F_{h,h'}$ and $\mathcal G_{h,h'}$. 

\medskip

We first analyze $I_1$: using the explicit expression of $\mathcal G_{h,h'}(\mb{k}',\mb{p};m)$ in \eqref{defmcGhh}, the properties of the dressed vertex functions mentioned after \eqref{ebruciavan}, the support properties of the single-scale propagator, as well as the dimensional bounds \eqref{boundsgh} on its derivatives, imply that the integral inside the summation over $h,h'$ in the definition of $I_1$ is bounded above by (const.)$2^{3h} |m|2^{-4h}$ (the factor $2^{3h}$ being proportional to the volume of the support of the integrand, while $|m|2^{-4h}$ is an upper bound on the integrand) times a characteristic function for the condition $|h-h'|\le 1$. Recalling that $m=\pm\epsilon$, we thus find that
\begin{equation} |I_1|\le C\epsilon\sum_{h\ge h_{3,\epsilon}}2^{-h}=C'\sqrt\epsilon,\end{equation}
where we used the definition of $h_{3,\epsilon}=\lfloor \log_2\sqrt\epsilon\rfloor$ in the identity. 

We next consider $I_2$: using the explicit expressions of $\mathcal G_{h,h'}(\mb{k}',\mb{p};m)$ and of $\mathcal F_{h,h'}(\mb{k}',\mb{p};m)$, the properties of the dressed vertex functions, the form of the single scale propagators $\hat g^{(h)}_{m}$ and $S_m$, their support properties and the bounds on their sup and on the sup of their derivatives,  as well as 
the bounds on $|\gamma_h-\gamma|$, on $|Z_{j,h}-Z_j|$ and on $|v_h-v|$, we find that 
the integral inside the summation over $h,h'$ in the definition of $I_2$ is bounded from above by (const.)$(2^h+|U|2^{\theta h})$. Recalling that the 
summand is non vanishing only if $|h-h'|\le 1$, we see that 
\begin{equation} |I_2|\le C\sum_{h=h_{1,\epsilon}}^{h_{2,\epsilon}}(2^{h}+|U|2^{\theta h})\le C'(\sqrt\epsilon+|U|\epsilon^{\theta/2}),\end{equation}
where we used the definition of $h_{2,\epsilon}={\lfloor \log_2(9\sqrt\epsilon)\rfloor}$ in the second bound. 

Finally, consider $I_3$. This term is a bit more delicate, because the number of terms involved in the sum is infinite and, therefore, we cannot a priori exchange the left derivative with the sum and then with the integral. We thus fix a $p_0<0$ small enough and derive a bound on the incremental ratio 
\begin{equation}\label{diagnondiag}\Delta_{p_0}\sum_{h,h'=-\infty}^{h_{2,\epsilon}}\int\limits_{\mathcal A_\epsilon}\! \frac{d\mb{k}'}{(2\pi)^3}\Big[\mc{G}_{h,h'}(\mb{k}',\mb{p};0)-\mc{F}_{h,h'}(\mb{k}',\mb{p};0)\Big],\end{equation}
where the action of $\Delta_{p_0}$ on a function $F(p_0)$ (recall that $\mb{p}=(p_0,\vec0)$) is defined as $\Delta_{p_0}F(p_0):=\frac1{p_0}\big[F(p_0)-F(0)\big]$. 
In \eqref{diagnondiag} we first focus on the contribution from the diagonal terms in the sum, i.e., those with $h'=h$. These can be written as:
\begin{equation}\begin{split} &\sum_{h=-\infty}^{h_{2,\epsilon}}\int\limits_{\mathcal A_\epsilon}\! \frac{d\mb{k}'}{(2\pi)^3}\int_0^1\!\! ds \,
\mbox{Tr}\Big\{\Gamma_{h,1}(\vec{k}',\vec{0})\hat{g}^{(h)}_{0}(\mb{k}')\Gamma_{h,2}(\vec{k}',\vec{0})\\
&\ \cdot \partial_{p_0}\hat{g}^{(h)}_{0}(\mb{k}'+s\mb{p}) 
+\gamma^2\sigma_2 S_{0}^{(h)}(\mb{k}')\sigma_1 \partial_{p_0}S_{0}^{(h)}(\mb{k}'+s\mb{p})\Big\},\end{split}\label{solodiag}\end{equation}
where $S^{(h)}_0(\mb{k})=f_h(\mb{k}')S_0(\mb{k}')$. Using the properties of the dressed vertex functions, the form of the single scale propagators $\hat g^{(h)}_{0}$ and $S_0^{(h)}$, their support properties and the bounds on their sup and on the sup of their derivatives, as well as the bounds on $|\gamma_h-\gamma|$, on $|Z_{j,h}-Z_j|$ and on $|v_h-v|$, we find that 
the integral inside the summation over $h$ in \eqref{solodiag} is bounded from above by (const.)$2^{3h}(2^{-2h}+|U|2^{-(3-\theta) h})$
(the factor $2^{3h}$ being proportional to the volume of the support of the integrand, while $(2^{-2h}+|U|2^{-(3-\theta) h})$ is an upper bound on the integrand). We thus find that 
\begin{equation}\big|\eqref{solodiag}\big|\le C\sum_{h\le h_{2,\epsilon}}(2^{h}+|U|2^{\theta h})\le C'(\sqrt\epsilon+|U|\epsilon^{\theta/2}),\label{boundsolodiag}\end{equation}
where we used the definition of $h_{2,\epsilon}$ in the second bound. 

We next focus on the contributions from the off-diagonal terms in \eqref{diagnondiag}, i.e., those with $h'\neq h$, such that either $\max\{h,h'\}<h_{2,\epsilon}$, or 
$\max\{h,h'\}=h_{2,\epsilon}$ and $\min\{h,h'\}\le h_{2,\epsilon}-2$ 
(note that for these terms, if $p_0$ is small enough, ${\text{supp}}(f_h)\cap {\text{supp}}(f_{h'}(\cdot+{\bf p}))$ is strictly contained in $\mathcal A_\epsilon$). After symmetrization under the exchange of $h$ with $h'$, 
these can be rewritten as
\begin{equation}\hskip-.2truecm\Big(\sum_{h'<h< h_{2,\epsilon}}\!\!\!+\sum_{\substack{h'\le h-2:\\ h=h_{2,\epsilon}}}\Big)
\int\! \frac{d\mb{k}'}{(2\pi)^3}\Big(I_{h',h}^{1,2}(\mb{k}',p_0)\ - I^{2,1}_{h',h}(\mb{k}',-p_0)\Big),
\label{rewsolonondiag}\end{equation}
where, recalling that $\mb{p}=(p_0,\vec0)$, 
\begin{equation*}\begin{split} &I_{h',h}^{i,j}(\mb{k}',p_0)=\int_0^1\!\! ds \,
\mbox{Tr}\Big\{\Gamma_{h,i}(\vec{k}',\vec{0})\hat{g}^{(h')}_{0}(\mb{k}')\Gamma_{h,j}(\vec{k}',\vec{0})\\
&\ \cdot \partial_{p_0}\hat{g}^{(h)}_{0}(\mb{k}'+s\mb{p}) 
+\gamma^2\sigma_j S_{0}^{(h')}(\mb{k}')\sigma_i \partial_{p_0}S_{0}^{(h)}(\mb{k}'+s\mb{p})\Big\}.\end{split}\label{solonondiag}\end{equation*}
By the same considerations spelled out after \eqref{solodiag}, we find that, for $h'<h$ and $(i,j)=(1,2), (2,1)$, 
$$\int d\mb{k'}\big|I^{i,j}_{h',h}(\mb{k}',p_0)\big|\le ({\rm const.})2^{3h'}\cdot2^{-h'}\cdot2^{-2h}\cdot(2^h+|U|2^{\theta h}),$$
so that 
\begin{equation}\begin{split}\big|\eqref{rewsolonondiag}\big|&\le 
 C\sum_{h'<h\le h_{2,\epsilon}}2^{2(h'-h)}(2^{h}+|U|2^{\theta h})\\ &\le C'(\sqrt\epsilon+|U|\epsilon^{\theta/2}),\end{split}\label{eq:65}\end{equation}
where we used the definition of $h_{2,\epsilon}$ in the second bound. We are left with the contributions from the terms with $\max\{h,h'\}=h_{2,\epsilon}$ and $\min\{h,h'\}=h_{2,\epsilon}-1$; by the same considerations as those used to bound the diagonal terms, these are bounded by (const.)$2^{h_{2,\epsilon}}+|U|2^{\theta h_{2,\epsilon}}$, which is of the same order as the right hand sides of \eqref{boundsolodiag} and \eqref{eq:65}.

In conclusion, $|I_3|\le 
C'(\sqrt\epsilon+|U|\epsilon^{\theta/2})$ and, therefore, the three terms $I_1,I_2,I_3$ all tend to zero as $\epsilon\to 0^+$. This  
proves \eqref{expbrack}, thus concluding the proof of the continuity 
bound \eqref{eq_17},  as desired. 

\section{Conclusion}

We studied the topological phase transition in an interacting version of the Haldane model, and proved that on the critical lines separating the normal and the topological insulating 
phases the transverse conductivity is quantized in semi-integer values, equal to the average of the integer values on either side of critical line. Together with 
the quantization results regarding the critical longitudinal conductivity proved in \cite{GJMP16} and the off-critical transversal conductivity proved in \cite{GMP17,GMP20}, our result proves 
the universality of the conductivity matrix in the whole topological 
phase diagram of the interacting Haldane model, provided the interaction strength is small enough compared to the bandwidth. 
Even though, for definiteness, we focus on a specific class of interacting lattice systems in this paper, 
we believe that our result applies more generally to the transition between distinct Hall phases of lattice interacting electron systems characterized at criticality by a semimetallic behavior, in the absence of on-site disorder. 

There are several open problems related to the results and methods introduced in this work, which deserve to be investigated. First, it would be interesting to consider the critical interacting Haldane model in a domain with boundary (say, in the half-plane) and investigate the nature of the edge theory: would it be possible to define and compute an edge Hall coefficient 
in such a critical, semimetallic, regime, matching the bulk value computed in this paper?

Similar questions, concerning both the bulk and edge transport coefficients, and the construction of a topological phase diagram analogous to the one in Fig.\ref{fig2}, can be asked for the thermal Hall conductance in the small temperature limit, see \cite{KS20}, where a Kubo-like formula for the bulk thermal conductance is proposed and related to its edge counterpart, which is supposedly given by the boundary chiral central charge.

Finally, it would be very nice to prove the universality of the longitudinal and transverse conductivities of critical (or quasi-critical) Hall systems in the presence 
of weak, marginally irrelevant, on-site disorder, in the spirit of \cite{LFSG94}, see also \cite{K83,Fr86a,Fr86b}. Marginally irrelevant contributions may in principle affect finite quantities at criticality such as the longitudinal and transverse conductivities (this is already the case for irrelevant contributions, such as those induced by lattice and short-range interaction effects, as discussed in this paper, and even more for marginally irrelevant ones): therefore, their effect deserves to be studied beyond the existing 
approximate RG or mean-field schemes. While a non-perturbative proof of universality in the disordered case seems beyond reach of the available rigorous methods, 
a systematic proof at all orders in renormalized perturbation theory may be within reach. We hope to come back to this and the aforementioned problems in future publications.

\bigskip

\acknowledgments

We thank Eduardo Fradkin, Marcello Porta, and Roberto Raimondi for useful comments, references and discussions. We gratefully acknowledge financial support from: the European Research Council through the ERC CoG UniCoSM, grant agreement n. 724939 (A.G. and R.R.) and the ERC-StG MaMBoQ, grant agreement n. 802901 (S.F.)}; MUR, PRIN 2022 project MaIQuFi cod. 20223J85K3 (A.G.); GNFM-INdAM Gruppo Nazionale per la Fisica Matematica. 

\bibliography{bibliography}

\end{document}